\documentclass[sn-apa]{sn-jnl}

\usepackage{amsthm}
\usepackage{epsfig}
\usepackage{amssymb}
\usepackage{bbding}
\usepackage{longtable}
\usepackage{multirow}
\usepackage{graphicx}
\usepackage{caption}
\usepackage{subfigure}
\usepackage{tabularx}
\usepackage{threeparttable}
\usepackage{color}
\usepackage{enumerate}

\jyear{2021}%

\theoremstyle{thmstyleone}%
\theoremstyle{thmstyletwo}%

\theoremstyle{thmstylethree}%

\begin{document}

\title[Research Status of Deep Learning Methods for Rumor Detection]{Research Status of Deep Learning Methods for Rumor Detection}


\author*[1]{\sur{Li Tan}}\email{tanli@th.btbu.edu.cn}

\author[1]{\sur{Ge Wang}}\email{wanggestu@163.com}

\author[1]{\sur{Feiyang Jia}}\email{jfy539@yeah.net}

\author[2]{\sur{Xiaofeng Lian}}\email{lianxf@th.btbu.edu.cn}

\affil*[1]{\orgdiv{School of Computer Science and Engineering}, \orgname{Beijing Technology and Business University}, \orgaddress{ \city{Beijing}, \postcode{100048}, \country{China}}}

\affil[2]{\orgdiv{School of Artificial Intelligence}, \orgname{Beijing Technology and Business University}, \orgaddress{ \city{Beijing}, \postcode{100048}, \country{China}}}


\abstract{{To manage the rumors in social media to reduce the harm of rumors in society. Many studies used methods of deep learning to detect rumors in open networks. To comprehensively sort out the research status of rumor detection from multiple perspectives, this paper analyzes the highly focused work from three perspectives: Feature Selection, Model Structure, and Research Methods. From the perspective of feature selection, we divide methods into content feature, social feature, and propagation structure feature of the rumors. Then, this work divides deep learning models of rumor detection into CNN, RNN, GNN, Transformer based on the model structure, which is convenient for comparison. Besides, this work summarizes 30 works into 7 rumor detection methods such as propagation trees, adversarial learning, cross-domain methods, multi-task learning, unsupervised and semi-supervised methods, based knowledge graph, and other methods for the first time. And compare the advantages of different methods to detect rumors. In addition, this review enumerate datasets available and discusses the potential issues and future work to help researchers advance the development of field.}}




\maketitle

\section{Introduction}\label{sec1}

{The Internet and social media have become comprehensive and large-scale platforms for disseminating real-time information. It is worth emphasizing that rumors can change the perceptions of billions of people during the rumor spread in social media. } According to the 2018 Internet trend report({\citet{meeker2018internet}}) ,{ more than a third of social media news events contain false information. The characteristic of rumors is spread widely and quickly. The latest research shows that the spread of rumors is 6-20 faster than non-rumors ~\citep{lazer2018science}. And the rumors get more people's attention by their emotional which cause spread widely ~\citep{zhao2015enquiring}. In addition, the spread of rumors also showed a lot of incredible harm. For example: in 2013, the Twitter account of the Associated Press was hacked. Then it post a claim that two explosions had occurred in the White House, and president to be injured. Although this rumor was debunked soon, it still spread to millions of users, causing serious social panic and leading to a rapid stock market crash ~\citep{intro1,domm2013false}. Even some rumors about the COVID-19 in 2020 pose a threat to life safety and increase the pressure on medical personnel, such as false statements suggesting drinking bleach to cure diseases ~\citep{intro2}.}

{Therefore, since 2011, researchers have made many efforts to establish a prioritized and automatic rumor detection method so that alleviate the pressure of rumors.} Figure 1 are representative works with the development of rumor detection. Among them, ~\citet{castillo2011information} conducted rumor detection related work on Twitter for the first time in 2011. They manually extracting  message-based features and topic-based features to be credible for platform-specific topic-related information degree to be evaluated. {In 2012, ~\citet{yang2012automatic} is the first research on rumor detection in Chinese social media. }In the same year,~\citet{gupta2012evaluating} used communication methods to solve the problem of rumors by constructing a network composed of users, messages, and events. {In 2013, \citet{kwon2013prominent} successfully extracted the temporal features of rumors for the first time. In 2015,  \citet{60} modeled the propagation mode of rumors in a propagation tree for the first time. In 2016, \citep{hoaxy} proposed the first rumor detection platform named Hoaxy, and for the first time mentioned the difference from propagation of rumors. However, there are hand-craft approaches above.} In the same year, ~\citet{ma2016detecting} learn the characteristics of the content of rumors by deep learning for the first time, opening a new page of rumor detection. {Then in 2017, \citet{18} applied CNN to rumor detection for the first time; \citet{19} first incorporated visual features to classify rumors. Their work leads to more multimodal fusion rumor detection work. In 2018, \citet{24} tried event-level cross-domain learning for the first time to solve the imbalance of rumors in different domains. \citet{85} first proposed a multi-task rumor detection method based on stance detection task. In 2019, \citet{35}. tried first to use the features of generative confrontation networks to simulate rumors. In 2020, the reinforcement learning method was first introduced in the rumor detection task by \citet{94}. In 2021, \citet{57} proposed dual emotional features, which is ahead of the existing emotional feature methods for rumor detection. The above methods are only part of the representative work of rumor detection.} Since deep learning has made incredible achievements in rumor detection, many researchers have extracted visual feature (~\citet{19,20}), propagation feature (~\citet{21,22}) and user social background feature (~\citet{chen2021catch}) of rumors from rich data. Based on the fusion of multiple features, more novel and excellent methods are proposed, such as multi-task methods~\citep{23}, adversarial learning~\citep{24}, semi-supervised methods~\citep{25}, weakly supervised methods~\citep{26,27} and so on.

{Due to the numerous researches related to rumor detection, it is difficult for new researchers to enter this field and grasp its current research status. Therefore, some surveys and reviews have conducted in-depth research on detecting rumors. For example, \citet{10} proposes an overview of how to develop a rumor detection system, which consists of four steps: rumor detection, rumor tracking, rumor stance classification, and rumor veracity classification. As their work pays less attention to feature extraction and neural network-based algorithms, it is impossible to summarize and compare different methods from an algorithmic point of view. The survey by \citet{sur1} described many methods based on manual features. However, due to the early publication, there is a lack of the latest deep learning methods. The survey by \citet{sur2} introduced the phased issues of the origin, dissemination, and detection of false information but also focused on machine learning methods, ignoring most of the deep learning methods. Overall, most of the published surveys focus on machine learning methods and only involve a small number of deep learning methods, which inspired our work. We aim to provide a comprehensive introduction to the rumor detection algorithm based on deep learning compared with the above survey. Based on deep learning, we focus on the three aspects of feature engineering, model structure, and algorithm perspective in rumor detection. In addition, we also analyzed the shortcomings of each algorithm. This analysis will help the future development of the algorithm.}

\begin{figure}[!t]
\centerline{\includegraphics[scale=0.47]{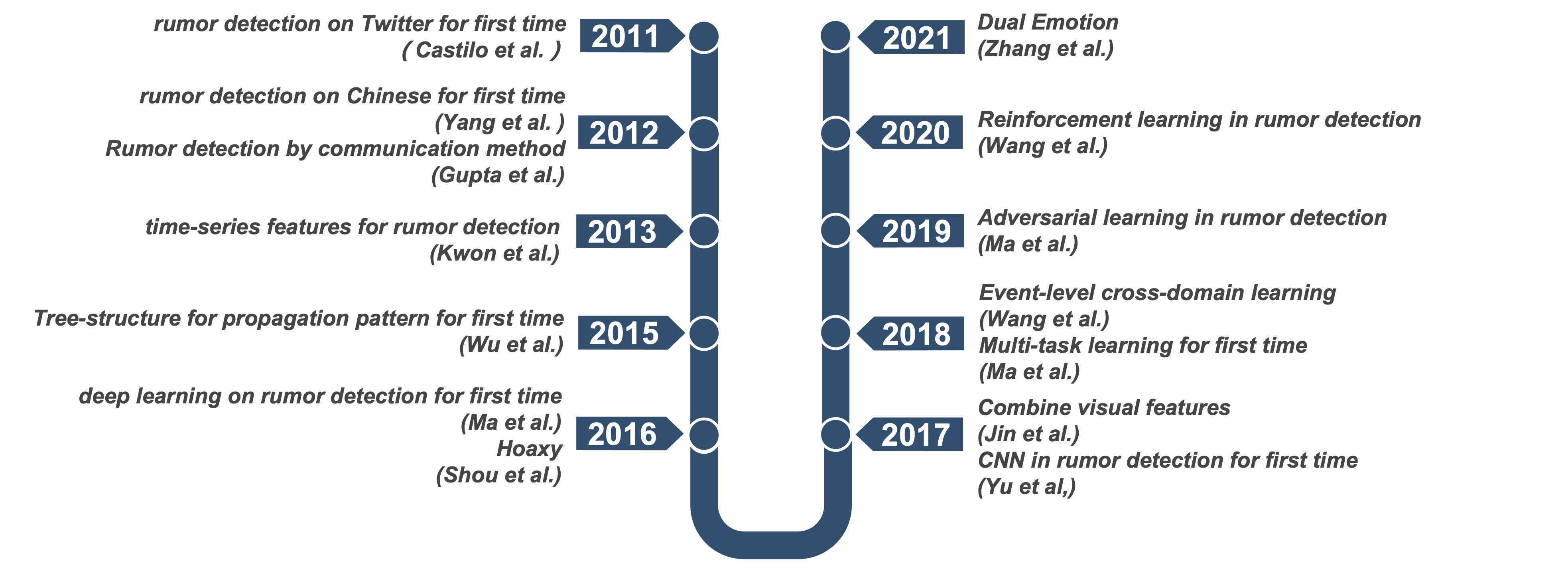}}
\caption{Development history of rumor detection\label{fig1}}
\end{figure}


\vspace{1em}

{The main contributions of the work are as under:
\begin{itemize}
	\item We discussed how to select and mine the feature of rumor from the massive data in the Internet, and how to use such feature in recent rumor detection research.
	\item As far as we know ,it is the first time for listing the rumor detection models based on deep learning and classify them into methods based on CNN, RNN, GNN, Transformer and discuss how to use each basic model and improved models in popular research to analyze the advantages and disadvantages.
	\item As far as we know, this article is the first to categorize popular research based on rumor detection methods, sort out and introduce the latest research ideas and methods.
	\item Summarize and describe the publicly available rumor detection datasets with size, label, number for each label, data type and remark.
\end{itemize}
}

\vspace{1em}

We will summarize the latest research and the most representative methods in rumor detection from three dimensions: characteristics, structure, and methods. The chapter structure of this article is as follows: {Chapter 2 introduced the review methodology of this article. }Chapter 3 introduces how to use data features to solve the problem of rumor detection in the latest research on rumor detection; Chapter 4 introduces the latest research models by model structure; Chapter 5 categorizes according to research methods Introduce each research content and method; in Chapter 6, we present the datasets available in rumor detection. Finally, in Chapter 7, we proposes potential challenges and issues to  existing research. Chapter 8 summarizes the work.

{\section{Research Methodology}}\label{sec2}

{\subsection{Development of The Review Protocol}}

{
To systematically investigate and sort out the research status of rumor detection, we conducted a systematic review of rumor detection based on deep learning. The first step of the review is to search for relevant research in multiple digital libraries and databases. Then the selection criteria are used to reduce the number of selected studies to further improve the quality of the papers involved and involve as many different deep learning methods as possible. After that, a set of research questions was formulated to thoroughly solve the research on the current situation of rumor detection.
}

{\subsection{Source of Information}}

{
To conduct our systematic literature review, the following digital libraries and datasets have been selected :
}

\vspace{1em}

\begin{enumerate}[-]
\item {IEEE Explore (www.ieeexplore.ieee.org)} 

\item {Springer Link (www.springerlink.com)} 

\item {Sciencez Direct (www.sciencedirect.com)} 

\item {ACM Digital Library (https://dl.acm.org)}

\item { Google Scholar (www.scholar.google.com)}
\end{enumerate}
\vspace{1em}

{The search keywords used to find related researchs are "Rumor detection'' OR ''Fake information'' AND ''Deep Learning'' OR ''Deep Neural Networks'' OR "[the name of Deep Learning Methods]". The searched literature results involve rumor detection and deep learning. To further improve the quality of the papers involved, we further screened the papers through multiple factors of the selection criteria. 
}




{\subsection{Research Questions}}

{
To sort out the comprehensive research status of rumor detection, we raised five research questions in Table 1. We sorted out and analyzed these five research questions in our recent research work and answered these five questions in turn through the subsequent chapters of this article.
}

\vspace{1em}

\begin{center}
\centering
\footnotesize
\begin{threeparttable}
\begin{tabular}{p{5cm}p{5cm}}

\multicolumn{2}{l}{\small{{\textbf{Table 1}}}}\\
\multicolumn{2}{l}{\small{{Questions considered in our systematic review.}}}\\

\hline
\label{table}\\

\textbf{{Question}}& \textbf{{Description}}\\

\hline
{
RQ1: What features of data mining rumors did these studies use and how to use?} & {This question lets researchers more quickly understand how to mine the features of rumors in massive data.}  \\
& \\

{RQ2: Which deep learning structures are used for rumor detection? }& {Researchers can understand more clearly which deep learning structures are more suitable for rumor detection problems and how to use them by this question.} \\
& \\

{RQ3: What deep learning methods are used in these studies? }& {This problem allows researchers to master the deep learning method of popular rumors detection faster. }\\
& \\

{RQ4: Which datasets are used mostly for conducting rumors analysis?} & {Identify available datasets helps researchers to use them as benchmarks as well as to compare with their work.}\\
& \\

{RQ5: What are the main challenges within the rumor detection field?} &
{The answer to this question helps new researchers recognize the open research challenges in this field.}\\

\hline

\end{tabular}

\end{threeparttable}
\end{center}

{\subsection{Selection Criteria}}

{The selection criteria are as follows:}

\vspace{1em}
\begin{enumerate}[i)]
\item {This article's rumor detection method for deep learning mainly covers the research work during 2017-2021. After deep learning has made fantastic progress in many fields, the iterative update speed of deep learning has dramatically increased. To enable researchers to grasp the current research status in this field more quickly, we focused on deep learning method research into nearly five years of work.} 

\item {To further improve the quality of the articles involved in this article, this article only screens high-quality conferences or journals (such as conferences and journals recommended by China Computer Federation ), or studies with high citation rates, or research groups that have made outstanding contributions to the field of rumor detection.} 

\item {This article only covers research papers written in English. } 

\end{enumerate}

\vspace{1em}
{In terms of screening, we eliminated researches irrelevant to the subject of the indexed literature and finally screened out 98 works with great attention by selection criteria and figure 2 is the distribution of the publication we selected.}

\begin{figure}[!htbp]
\centering  
\subfigure[Year-wise]{   
\begin{minipage}{5cm}
\centering    
\includegraphics[scale=0.3]{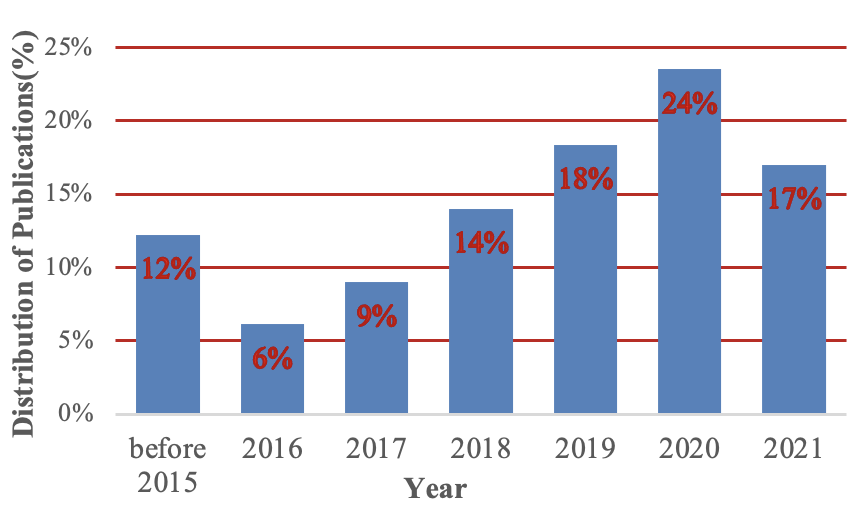}  
\end{minipage}
}
\subfigure[Topic-wise]{ 
\begin{minipage}{6cm}
\centering    
\includegraphics[scale=0.417]{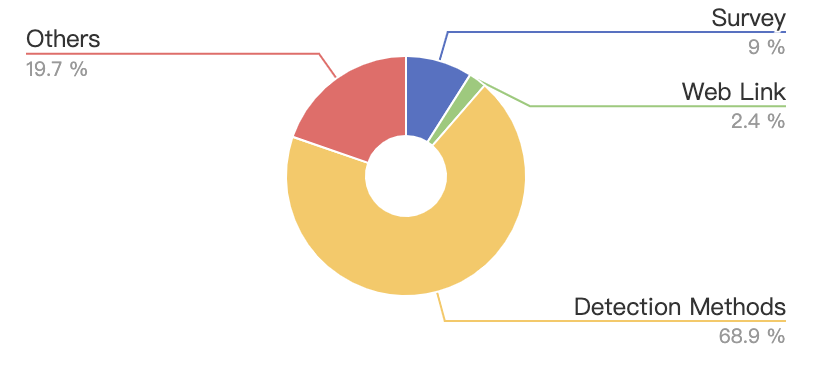}
\end{minipage}
}
\caption{Distribution of publications.}    
\label{fig:1}    
\end{figure}





\section{Feature Extraction}\label{sec3}

Rumor detection will be regarded as a binary classification problem in most cases, and a few studies will treat it as a multi-classification problem. Most of the literature follows the general learning rules of supervised classification in machine learning: first, extract the characteristics of the representative samples from the two types of samples; second, train the extracted samples to a suitable model; finally, test or test the data sets with unknown labels. It is to evaluate its label. In the latest literature, most of the research focuses on how to extract features and where to extract features. How to deal with the characteristics after obtaining the characteristics is also a critical issue of the research. {To answer RQ1, in this chapter, we review the content features, social features and propagation structure features used in rumor detection, and introduce the methods of feature extraction and how to use features.}

In social media, general news or blog content contains a wealth of multimodal resources, such as pictures, videos, texts, etc., as well as some links, comments, likes, and other information associated with the communication process. As shown in Figure 3.  In addition, in the latest literature, it is found that there is a big difference in propagation structure between rumors and real information. Therefore, there is a lot of work to extract features from the propagation structure.

\begin{figure}[!h]
\centerline{\includegraphics[scale=0.5]{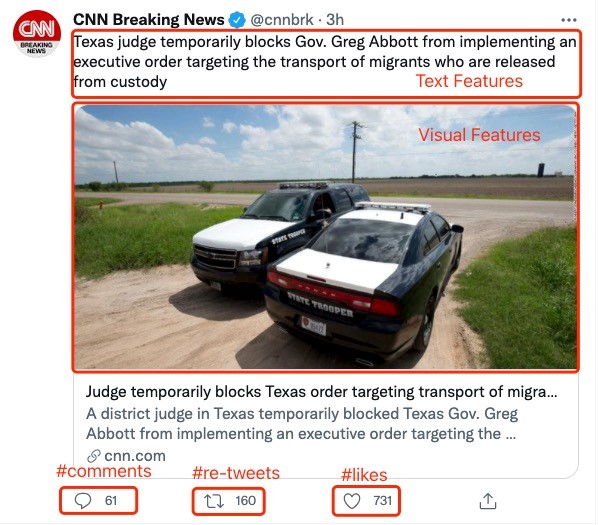}}
\caption{ Example of Twitter\label{fig2}}
\end{figure}

\subsection{Content Feature}
\subsubsection{Text Features}

Compared with real information, rumors are used to confuse and mislead the public. The text makes people more emotional and can attract more people's attention. Traditional text features are mainly divided into three aspects: lexical features, syntactic features and topic features. Lexical features refer to language features extracted from a single word or a single word level. In the early detection of rumors with manual features, ~\citet{castillo2011information} calculated the "number of words", "number of words", "number of different words", and other indicators based on the words contained in a message. Syntactic feature refers to a type of feature that is sobbing from the sentence level, such as part-of-speech tagging, word frequency, etc. The topic feature refers to the text feature extracted from the entire message collection, and the purpose is to understand the message and its implicit semantics.

Unlike manually extracting features from the text in the past, most latest studies(~\citet{ma2016detecting,18,31,32,33,34,35,36,37})  no longer only use text feature, but the fusion of multiple text features solves the problem of rumor detection. For example: ~\citet{ma2016detecting} introduced a recurrent neural network to learn hidden representations from the text content of related articles, opening a precedent for using deep learning to automatically extract the features of rumor texts. Later,~\citet{18}  used convolutional neural networks to obtain key features and hidden semantic features from text content. However, these studies take the full text as input only and lose the article's original chapter structure. So, ~\citet{31}  adopted the hierarchical structure of "word-sentence-article" to understand a text and extracted the word level and sentence hierarchy as the feature of understanding the text. ~\citet{37}  learned the word-level and sentence-level representation of statement texts and news articles as text features. However, the above research did not consider the connection between the article and terminology, \citet{24,26} believe the terminology of different domains will affect the effect of rumor detection, so they used terms to distinguish fake news and capture the potential relationship between articles and terms.

Studies have shown that most fake news and rumors have much emotional information, and the topic of rumors generally contains words that attract attention, which is used to drive readers’ emotions and attract readers to read to promote spread. Therefore, the latest research uses the emotional information in the article as one of the characteristics of judging rumors. ~\citet{32} manually extracted words with emotional characteristics such as emotion, morality, and exaggeration words in the news, combines them with theme characteristics to obtain fake news and real information in emotions and topics.~\citet{57} used the emotional information of news content in the experiment and extracted the emotional information from news reviews and used dual emotional features to detect fake news. ~\citet{33} used LDA (Latent Dirichlet Allocation) to extract topic features combined with text features extracted by XLNet{\citep{XLNet}} as critical features.

\subsubsection{Visual Features}

Even if the method using text features in rumor detection has achieved good results, in general, rumors have multi-modal information, such as images, videos, etc. Many studies combine visual feature to the features of rumor detection. During the hand-craft feature period, visual features can be divided into visual statistical features and visual content features. Among them, visual statistical feature is the statistics of rumor image, such as the number of images or the propagation time of the false image, and so on. The visual content feature refers to the content in the visual image, such as the clarity and diversity of the visual image. The early research of ~\citet{38} was the first attempt to manually extract information such as the propagation time of fake images on Twitter and proposed a classification model to identify fake images on Twitter. In the latest literature, many studies prove the importance of visual information (~\citet{39,40}). ~\citet{40}pointed out that images are more influential than texts and often appear in rumors.

In recent years, most of the research directly extracts high-dimensional representations of images from visual information (such as images and videos) using deep learning pre-training models for calculation. Such as research (~\citet{24,31,43,45}) extract fake news, rumor and tweets image information by pre-trained convolutional neural network(CNN), mining deep-level visual features, and extract visual information High-dimensional representation is combined with text and other modal features: ~\citet{44} used residual networks to learn visual features in fake news, and for the first time proposed visual separation representation learning, which can remove features in specific areas of visual features, so that model learns cross-domain features to realize cross-domain event fake news detection.

In addition, there are studies using image embedding methods to convert images into matrices. ~\citet{151} embed visual information into a matrix by image2sentence (~\citet{47})  and then used TextCNN{(Text Convolutional Neural Network,\citet{TextCNN} ) } to extract vision features. Compared with the pre-trained model using convolutional neural networks such as VGG({Visual Geometry Group}, \citet{VGG}), it can calculate the similarity between different modal data and increase the receptive field.

There are also studies using deep learning methods to extract the content information contained in images. For example, ~\citet{40} used the research results of ~\citet{41} to extract image emotional information in pictures.

\subsection{Social Feature}
In social media, a series of social interactions occur during rumors propagate. Under normal circumstances, interactions in social media are divided into three categories: follow, link, and forward comments. Social background information will be generated during the interaction process, and research has found that users with special characteristics in the user group participating in rumor interaction, such as low-credit users such as "marketing account", will forward and comment on rumors that are sufficiently eye-catching. To increase their exposure to the public. This survey divides the social background information into user feature and propagation feature.

\subsubsection{User Feature}
User feature are derived from the user's social network. Rumors are created by a few users and spread by many users due to profit factors. The analysis of user feature can provide critical clues for rumor detection. User feature include individual features and group features. Among them, individual features are extracted from a single user, such as: "registration time", "age", "identity authentication", etc. In the early research, the reliability of users was evaluated from the perspective of reporters, and information such as "user credibility" and "user location" were used as user characteristics. Features of the user group are the characteristics extracted from the user group, such as "verified user ratio" and so on.

~\citet{48} modelled user participation by selecting social concerns of critical user comments and set 22 social characteristics, including "proportion of users with avatars", "proportion of users verified by verification", and "average number of fans". Among them, there are eight user characteristics. ~\citet{49} extracted user credit features to represent users' social reactions.~\citet{23} combined the user's credit information with the user's comment information on the statement to solve the task of rumor detection. ~\citet{50} extracted user personal information in the retweeted user sequence as user features.

\subsubsection{Propagation Feature}
In the process of dissemination, rumors will be forwarded, commented, and liked by most people. Propagation features include features extracted from some information that appears in the process of communication, such as: "user comments", "number of reposts", "number of likes", "number of clicks", and so on.

Some studies extract features from user comments to determine users' attitudes toward rumors. For example, ~\citet{49} proposed mining users' conflicting views from social media posts and estimating their credibility values to detect fake news. There is also much work (\citet{23,51,54}) extracting position information based on users' replies to posts and statements. Most of these works use users' replies to statements as propagation information for position detection tasks as the key to completing auxiliary rumor detection tasks.

In some studies, user comments collected during the dissemination process are extracted and combined with text features to solve rumors. ~\citet{54} build heterogeneous graphs based on posts on social networks, using comments and related users. ~\citet{52} uses social contextual content (such as comment replies, forwarding replies) and social context metadata (such as the number of reposts, the number of likes, and the number of favorites) for modeling. In research of \citet{55},they extract semantic information from users' comments on rumors, find consistent comments, and capture the implicit relationship between comments and rumors to enhance the model's semantic reasoning ability further. However, the above research only considers the connection between users and rumors and ignores the relationship between users. Therefore, ~\citet{53} correlated the conversation threads of comments in pairs to extract the hidden dependencies between user interactions.  The user-to-user mutual attention method simulates the potential influence relationship between users in the propagation path to stably capture the mode and connotation related to the spread of rumors and non-rumors.(~\citet{58})

\subsection{Propagation Structure Feature}

The propagation structure feature records and reproduces the propagation process of rumors and uses the user's reposting as a node to construct a reposting graph. Studies have pointed out that rumors and non-rumors are significantly different in the propagation structure (~\citet{59}). Figure 4 shows the comparison of the dissemination structure of non-rumors and rumors. The communication structure features include the relationship or similarity between articles, the relationship between users, the feature of the communication path, etc. The key communication features for judging rumors are extracted through the communication path and the communication graph. The earliest method of using propagation structure features to apply to rumor detection (~\citet{gupta2012evaluating}) formed a network of users, messages, and events, and unilaterally assumed that trusted users would not provide credibility for rumor events and that the links to trusted messages were better than rumor .The weight of information is more significant. However, some trusted users might be confused by disguised rumors, resulting in unsatisfactory results.

\begin{figure}
\centerline{\includegraphics[scale=0.8]{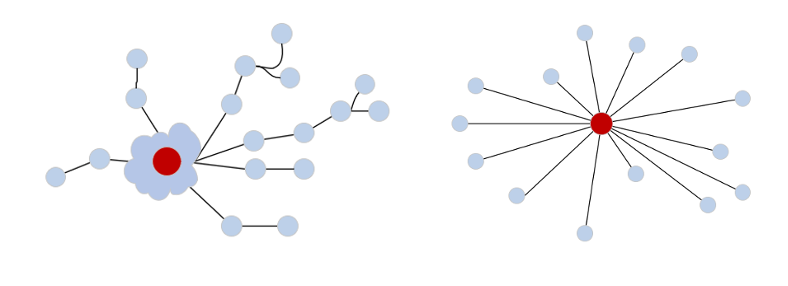}}
\caption{Example of non-rumor spreading (left) and a rumor spreading example (right)\label{fig4}}
\end{figure}

Some studies have proposed the use of graph kernels to capture communication features. For example, ~\citet{60} proposed a hybrid SVM (Support Vector Machine) classifier, combined with RBF (Radial Basis Function), based on the random walk graph kernel to capture the feature of propagation. Some studies model the propagation path of rumours as a propagation tree. For example, ~\citet{22} first proposed the concept of propagation tree core in the rumor detection on Twitter, starting with the propagation mode and using core learning. The tree models Twitter and compares the nuclear learning tree of non-rumors and rumors to analyze whether it is a rumor.

In addition, some studies regard the spread of rumors as a sequence with time-series feature. In the study of ~\citet{61}, media users and social network structures are embedded, and then LSTM (Long Short Time Memory) is used. To represent and classify the propagation path of the message, however, the study did not consider the time sequence change of the diffusion structure path. In the latest research, ~\citet{68} proposed to participate in changes along the path of the diffusion structure and the time of response to integrate the feature of the propagation structure. The above research still focuses on extracting the propagation structure's feature through the propagation sequence, but the propagation sequence may lose key structural information (~\citet{63}). Therefore ~\citet{63} uses Node2Vec{\citep{node2vec}} method, embeds the nodes in the propagation graph as a matrix as the communication information, and uses the convolutional layer to extract the propagation features. However, Node2Vec does not consider the weights in the figure, and in social media networks, accounts with different reputations have different propagation path weights. So more work began to join the Graph Neural Network (GNN) method to extract the propagation structure features from the adjacency matrix. 

In the latest research, many works have proved that graph neural networks can cope with the characteristics of the spreading structure of rumors well(~\citet{64,65,66,67}). For example: ~\citet{65} used graph neural networks to capture the similarities between disseminated fake news. In the same year, ~\citet{66} used graph convolutional networks(GCN) to locate multiple rumor sources without prior knowledge of the propagation model, and used the multi-level neighborhood information of nodes to establish node representations to improve the accuracy of the source prediction. ~\citet{67} used bottom-up and top-down GCN to extract the spread and spread patterns of rumors respectively.

~\\

\section{Model Structure}\label{sec4}

The second chapter introduces the features and extraction methods currently used in rumor detection according to feature dimensions.{To answer RQ2, we will introduce the rumor detection method in the dimension of the structure.}

\subsection{Rumor Detection Based on CNN}

The convolutional neural network(CNN) is a feedforward neural network that includes convolutional calculations and has a deep structure. It is one of the representative algorithms of deep learning.{It and its improved models (~\citet{110}) have been proven to perform well in computer vision (~\citet{69}), natural language processing (~\citet{70}) and other fields.} Table 2 shows some representative examples of using CNN for rumor detection tasks in the latest research and details the latest method.

In recent years, most studies (~\citet{18,31,62,71}) use convolutional neural networks to extract hidden features of text content and visual information. For example: ~\citet{18} first used convolutional neural networks for rumor detection tasks and used convolutional neural networks to extract key features from the text content of related posts for rumor detection. ~\citet{71} proposed an attention residual network, which can obtain important information in local and global content features by learning text content information in rumors. Use this to understand the classification of rumors. In addition, some research (~\citet{24}) uses CNN to extract the visual features of fake news combined with text features.

In the latest research, CNN has been used to extract the features of the embedding matrix in many researches. For example, ~\citet{151} used fake news text and visual content to embed it and then used a convolutional neural network to extract it as a feature map to calculate cross-modal similarity to judge the inconsistency between the text and the image description. Similarly, ~\citet{63} used convolutional layers in the proposed Rumor2vec framework for feature extraction after text embedding and node embedding. They combined the two features to make the framework learn a joint representation of text and propagation features to detect rumors.

Many studies have proved the effectiveness of convolutional neural networks in extracting rumor features, which can obtain local key features well. However, in convolution, the timing information of the input vector is ignored and disrupted. However, whether it is the text data of the rumor or the spread sequence of the rumor, the timing information can play a vital role in rumor detection.

\begin{center}
\centering
\footnotesize
\begin{threeparttable}
\begin{tabular}{p{0.8cm}p{0.8cm}ccccp{1.4cm}p{1.8cm}p{2cm}}

\multicolumn{9}{l}{\small{\textbf{Table 2}}}\\
\multicolumn{9}{l}{\small{Detail of rumor detection model based on CNN.}}\\

\hline
\label{table}\\

\textbf{Paper}&\textbf{DL Model}&\multicolumn{4}{@{}@{}c@{}@{}}{\textbf{Features in experiment}} & \multicolumn{1}{@{}c@{}}{\textbf{Dataset}} & \multicolumn{1}{c@{}}{\textbf{Description}} & \multicolumn{1}{@{}c@{}}{\textbf{Characteristic}}\\
& & \textbf{T}  & \multicolumn{1}{@{}c@{}}{\textbf{S}}  & \textbf{V} & \textbf{P S}& &&  \\
\hline 

 \citet{31} & CNN  & \Checkmark  & \XSolidBrush & \Checkmark  & \XSolidBrush  & Twitter & CNN extract text and visual features & Combining displayed features and hidden features to jointly judge fake news   \\

 \citet{71} & CNN  & \Checkmark   & \XSolidBrush & \XSolidBrush& \XSolidBrush & Twitter &Combination of residual unit and attention module & Attention residual network can capture long-distance information and mine deep correlation   \\

 \citet{151} & CNN  & \Checkmark   & \Checkmark & \XSolidBrush& \XSolidBrush & PolitiFact GossipCop  & Extracting text and visual feature using Text-CNN  & The semantic consistency of text and image is detected. Image embedding using image2sensence.   \\

 \citet{63} & CNN  & \Checkmark  & \XSolidBrush& \XSolidBrush  & \Checkmark & Twitter15 Twitter16  & Extracting text features and propagation structure features from 1D CNN & The propagation structure is embedded into a vector. Connecting text features and communication structure features  \\

 \citet{32}& CNN, RNN & \Checkmark   & \XSolidBrush & \XSolidBrush& \XSolidBrush & MutiSource-Fake & CNN and Bi-GRU are combined to extract text and emotional feature respectively & CNN and Bi-GRU are combined to extract text   \\

 \citet{24}& CNN, RNN & \Checkmark   & \XSolidBrush & \Checkmark & \XSolidBrush & Twitter Weibo & TextCNN extracts text features and VGG extracts visual feature & Multi-modal feature fusion, using adversarial learning to obtain cross-domain features  \\

\hline
\end{tabular}
\begin{tablenotes}
\footnotesize
\item DL:Deep Learning; T:Text; S:Social; V:Vision; P S:Propagation Structure
\end{tablenotes}
\end{threeparttable}
\end{center}

~\\

\subsection{Rumor Detection Based on RNN}

Recurrent Neural Network (RNN) with a sequence of data as input is recursive in the sequence's evolution direction, and all nodes are connected in a chain. The most significant difference between it and the convolutional neural network is that it retains the last state during recursion and retains timing information. Therefore, most existing researches are keen on using recurrent neural networks for natural language processing (~\citet{72}). ~\citet{ma2016detecting} first cited RNN for the rumor detection task, automatically learning Twitter content based on time series, using TF-IDF to model words, and then using RNN to learn the potential content of rumors. Since then, recurrent neural networks and their variants have been used more in rumor detection, and many of them are the latest research results.

Most studies (~\citet{19,32,43,73,75}) use RNN to learn text features, image features, and multimodal fusion features. For example: ~\citet{43} used multiple stacked two-way LSTMs to learn text features and used LSTMs to fuse the text and social background features. ~\citet{56} used RNN to obtain word-level representations of news content, sentence-level representations of news content, and user comment representations to describe the modeling from news language features to potential feature spaces. ~\citet{73} proposed a BiLSTM with multiple loss levels based on attenuation factors to deal with rumors detection on Twitter. These two directions can extract deep context information from a limited amount of text.

In addition, there are also studies using the characteristics of RNN to retain timing to capture the timing characteristics of rumors in propagation. ~\citet{21} modeled the propagation structure of rumors as a tree structure and used GRU{(Gate Recurrent Unit)} to calculate each branch of the tree sequence. ~\citet{77} proposed an LSTM tree based on a convolutional unit to predict the authenticity of positions and rumors in social media conversations, using the source blog post as the root node and the response as the child node. In each node, LSTM is used. To learn the position of evidence or comment, to carry out position detection. In addition, ~\citet{61} used LSTM to represent and classify the propagation path of messages, which was used to solve malicious spreaders disguising fake news as real news. In the study, the news propagation tree structure in research of ~\citet{21} was converted into time. At the same time, LSTM can be used to use the dependency between items separated by a long distance in the sequence.

Some studies (~\citet{23,74}) use LSTM as a shared unit in multi-task learning. They use LSTM to share the hidden vectors calculated by multiple meta-tasks to assist in rumor detection. For example, ~\citet{23} and ~\citet{74} use LSTM to fuse position detection and rumor detection as two meta-tasks.

Although RNN can extract timing information, the input of RNN is generally a sequence, which leads to the destruction of the propagated structural information, and the effect of extracting the propagated structural features is not good.

~\\
\vspace{3em}

\begin{center}
\centering
\footnotesize

\begin{longtable}{p{0.8cm}p{0.7cm}ccccp{1.3cm}p{2.1cm}p{2.3cm}}

\multicolumn{9}{l}{\small{\textbf{Table 3}}}\\
\multicolumn{9}{l}{\small{Detail of rumor detection model based on RNN.}}\\

\hline
\label{table}\\

\textbf{Paper}&\textbf{DL Model}&\multicolumn{4}{@{}@{}c@{}@{}}{\textbf{Features in experiment}} & \multicolumn{1}{@{}c@{}}{\textbf{Dataset}} & \multicolumn{1}{c@{}}{\textbf{Description}} & \multicolumn{1}{@{}c@{}}{\textbf{Characteristic}}\\
& & \textbf{T}  & \multicolumn{1}{@{}c@{}}{\textbf{S}}  & \textbf{V} & \textbf{P S}& &&  \\
\hline 

  \citet{43} & RNN  & \Checkmark  & \XSolidBrush & \Checkmark  & \XSolidBrush  & Twitter15 Twitter16 & Bi-LSTM extracts text features & Encoder-decoder structure,end-to-end framework\\

 \citet{75} & RNN  & \Checkmark   & \Checkmark  & \Checkmark & \XSolidBrush & Twitter Weibo &LSTM combined with Attention layer & Fusion of visual features, social features, and text features \\

 \citet{73} & RNN  & \Checkmark   & \XSolidBrush  & \XSolidBrush & \XSolidBrush & PHEME &Bi-LSTM extracts text features & Divided into blog post level and event level modules, event level modules increase attenuation factor\\

 \citet{21} & RNN  & \Checkmark   & \XSolidBrush  & \XSolidBrush & \Checkmark & Twitter15 Twitter16 &GRU as a hidden unit &Extract features along the propagation tree and preserve the propagation structure\\

 \citet{77} & RNN  & \Checkmark    & \Checkmark & \XSolidBrush  & \XSolidBrush & PHEME &LSTM extracts text content and comment content features & LSTM-Tree structure, keeping the relationship between source posts and replies\\

 \citet{61}& RNN   & \XSolidBrush & \Checkmark & \XSolidBrush & \Checkmark & Twitter &LSTM extracts propagation structure features &Keep sequence and get farer dependencies\\

 \citet{23}& RNN   & \Checkmark & \Checkmark & \XSolidBrush & \XSolidBrush & RumorEval PHEME &LSTM to extract user comments and fuse two task features & user characteristics to position detection tasks\\

 \citet{74}& RNN   & \Checkmark & \Checkmark & \XSolidBrush & \XSolidBrush & RumorEval &GRU extracts features after fusion of text and visual information &Better integration of two different types of meta-tasks\\

\citet{56}&RNN & \Checkmark & \Checkmark & \XSolidBrush & \XSolidBrush &FakeNewsNet &Bi-GRU extracts the contextual semantics of text content and user comments &Explore interpretable information from user reviews\\

 \citet{35}&RNN & \Checkmark & \XSolidBrush  & \XSolidBrush & \XSolidBrush &Twitter PHEME &GRU storage hidden repress-entation & Generative adversarial learning, divided into GRU encoder and GRU decoder\\

 \citet{57}&RNN & \Checkmark & \Checkmark  & \XSolidBrush & \XSolidBrush &RumorEval Weibo16 Weibo20& Bi-GRU extracts original content features and emotional features in comments & Combine publisher sentiment and social comment sentiment\\

 \citet{50} &RNN & \Checkmark & \Checkmark  & \XSolidBrush & \XSolidBrush  & Twitter15 Twitter16 &GRU extracts the propagation process and original text features &GCN is used as graph perception, CNN and GRU jointly extract propagation features\\

\hline
\end{longtable}
\begin{tablenotes}
\footnotesize
\item DL:Deep Learning; T:Text; S:Social; V:Vision; P S:Propagation Structure
\end{tablenotes}
\end{center}

\subsection{Rumor Detection Method Based on GNN}

In recent years, graph neural networks have received a lot of attention. The social network is a common type of graph data representing the social relationship between various individuals or organizations. At the same time, the graph neural network uses non-Euclidean graphs as input, so compared to CNN and RNN, it can Keep the structure of the spread of rumors. The latest research shows that the dissemination structure of rumors and true information is different (~\citet{59}).

In the latest work, GNN is mainly used to extract the structural features of propagation or the structural features of user interaction. ~\citet{76} used stacked graph convolutional neural network layers to extract the structural feature of dissemination from heterogeneous graphs composed of various data such as users, user comments, news dissemination, etc., to detect rumors automatically. ~\citet{67} proposed using a two-way graph convolutional network in their latest work, using top-down and bottom-up propagation directions to simulate the spread and spread of rumors. In addition, ~\citet{65} also used a graph neural network to propose a semi-supervised fake news detection method to solve training in a limited number of labeled articles.

~\citet{lin2020graph} proposed a rumor detection method based on a graph autoencoder, which uses an encoder to use an efficient graph convolutional network to treat the initial text and the propagation graph as input and update the representation vector through propagation to learn the text and dissemination of information.

~\citet{50} proposed a social media rumor detection method based on graph structure adversarial learning to deal with rumors in social networks disguised in various ways to avoid rumor detectors. This study established a heterogeneous information network to simulate the rich information between users, posts, and comments using a graph confrontation learning framework. The encoder tries to dynamically add intentional disturbances to the graph structure to trick the detector, containing disguised heterogeneous graphs. Input to the graph convolutional neural network to obtain the perturbation structure representation of each blog post. The detector will learn more unique structural features to resist this perturbation, thereby enhancing the detector's camouflage way to debunk rumors, and learn the characteristics of diversified patterns.

~\citet{54} mainly focused on integrating complex semantic information by learning forwarding sequences and how to model the heterogeneous graph structure of all microblogs and participants globally for rumor detection. Research uses graph attention network to capture global semantic information in heterogeneous graphs constructed by posts, comments and related users on social networks. The research uses a graph convolutional neural network to generate the latent representation of each node, uses the attention mechanism to obtain the weight of each node and iterates, and finally captures the global representation.

Table 4 is the detailed information of the latest GNN-based rumor detection work. From these studies, it can be found that the GNN-based multi-modal rumor detection data is not well used. This may also be one of the research directions of future work.

\begin{center}

\begin{table}[!htbp]
\centering
\footnotesize
\begin{tabular}{p{1cm}p{0.8cm}ccccp{1.4cm}p{1.8cm}p{2cm}}

\multicolumn{9}{l}{\small{\textbf{Table 4}}}\\
\multicolumn{9}{l}{\small{Detail of rumor detection model based on GNN.}}\\

\hline
\label{table}\\

\textbf{Paper}&\textbf{DL Model}&\multicolumn{4}{@{}@{}c@{}@{}}{\textbf{Features in experiment}} & \multicolumn{1}{@{}c@{}}{\textbf{Dataset}} & \multicolumn{1}{c@{}}{\textbf{Description}} & \multicolumn{1}{@{}c@{}}{\textbf{Characteristic}}\\
& & \textbf{T}  & \multicolumn{1}{@{}c@{}}{\textbf{S}}  & \textbf{V} & \textbf{P S}& &&  \\
\hline 
\citet{76} & GNN  & \XSolidBrush  & \XSolidBrush  & \XSolidBrush & \Checkmark &unnamed & Four layers of GCN and two convolutional layers & Combine user profiles, activities, networks and communications, and content\\

\citet{67} &GNN& \XSolidBrush  & \XSolidBrush  & \XSolidBrush & \Checkmark &Twitter15 Twitter16 &GCN extracts propagation structure features &Bi-GCN extract propagateion and diffusion features from top-down and bottom-up respectively\\

\citet{50} &GNN CNN RNN & \Checkmark & \Checkmark & \XSolidBrush & \XSolidBrush &Twitter15 Twitter16 &GCN extracts propagation feature &Graph structure adversarial learning\\

\citet{65} & GNN & \Checkmark & \XSolidBrush & \XSolidBrush & \XSolidBrush & \citet{61} &Attention Graph Neural Network to extract graph features compose of text vectors&Semi-supervised learning method\\

\citet{lin2020graph} & GNN &\Checkmark  & \XSolidBrush  & \XSolidBrush & \Checkmark &Twitter15 Twitter16 &GCN extracts propagation structure features in the encoder &Graph Auto-encoder Extract both dissemination structure features and text structure features at the same time\\

\citet{54} &GNN CNN &\XSolidBrush  & \XSolidBrush  & \XSolidBrush & \Checkmark &Twitter15 Twitter16 Weibo & GCN extracts graph structure features &Graph attention network to obtain structural semantics\\

\citet{66} &GNN &\XSolidBrush  & \XSolidBrush  & \XSolidBrush & \Checkmark &Twitter15 Twitter16 &GCN extracts graph structure features&Use the characteristics of transmission structure to judge the accuracy of the source of rumors\\

\citet{68} &GNN RNN &\Checkmark  & \XSolidBrush  & \XSolidBrush & \Checkmark &PHEME    RumorEval &GCN extracts nonlinear diffusion feature &Combination of dual propagation feature ,GCN extracts nonlinear features, LSTM extracts time series linear features\\

\hline
\end{tabular}
\begin{tablenotes}
\item DL:Deep Learning; T:Text; S:Social; V:Vision; P S:Propagation Structure
\end{tablenotes}
\end{table}
\end{center}

\subsection{Rumor Detection Based on Transformer}

Since CNN is easy to parallelize, it is not suitable for capturing dependencies within the field sequence. RNN can capture the dependence of long-distance sequences. However, it is challenging to realize parallel processing sequences. In order to integrate the advantages of CNN and RNN, ~\citet{80} combined the attention mechanism to design Transformer. This model uses the attention mechanism to achieve parallel capture of sequence dependencies and can process tokens at each position of the sequence at the same time. In recent years, Transformer has been proven to perform well in natural language processing tasks such as machine translation (~\citet{81}). Therefore, there are studies to migrate the Transformer structure to the rumor detection work.

Since Transformer can capture the dependence of long-distance field sequences, ~\citet{53} used the multi-attention mechanism in Transformer to model the long-distance interaction between tweets. The tree structure information is disturbed in the research, and the dependence is obtained from user comments between different conversation threads. Use Transformer's multi-attention mechanism to obtain the characteristics of each original blog post and retweet comment, calculate the correlation between them and give the correlation weight so that the user's comment characteristics can be used as the basis for judging rumors more accurately.

~\citet{82} used the establishment of a transformer specific to position detection and a cross-task Transformer. They used the multi-head attention mechanism to guide the model to pay more attention to the particular characteristics of position rumors, to obtain the dependence of position and rumors, and respectively For rumor detection, use predicted position tags as the basis for judgment.

Transformer is based on the encoder-decoder architecture. The Transformer's encoder can encode word embedding vectors into high-dimensional representations containing text features. Therefore, some studies (~\citet{83,86}) use Transformer's encoder to extract features in word vectors. For example, ~\citet{83} and ~\citet{86} encoded all news through the Transformer encoder to generate the representation of the news and extract text features for subsequent calculations.

~\citet{87} proposed a Transformer tree-based method to utilize user interaction in conversations. The study models the propagation of each statement as a tree structure. Transformer is established as a bottom-up Transformer, a top-down Transformer, and a hybrid Transformer model in this work. The former compares the response tweets of each book borrowing point in pairs to capture the consistent attitude towards each tree node. The latter describes how information flows from the source blog post to the current node. Finally, a hybrid Transformer is used to fuse position features and structural features for rumor detection.

Table 5 is the detailed information of the latest work based on Transformer's rumor detection. The existing Transformer-based rumor detection work does not make good use of the dissemination of structural information.

\begin{center}

\begin{table}[!h]
\centering
\footnotesize
\begin{tabular}{p{1cm}p{1.2cm}ccccp{1.4cm}p{1.8cm}p{1.8cm}}

\multicolumn{9}{l}{\small{\textbf{Table 5}}}\\
\multicolumn{9}{l}{\small{Detail of rumor detection model based on Transformer.}}\\

\hline
\label{table}\\

\textbf{Paper}&\textbf{DL Model}&\multicolumn{4}{@{}@{}c@{}@{}}{\textbf{Features in experiment}} & \multicolumn{1}{@{}c@{}}{\textbf{Dataset}} & \multicolumn{1}{c@{}}{\textbf{Description}} & \multicolumn{1}{@{}c@{}}{\textbf{Characteristic}}\\
& & \textbf{T}  & \multicolumn{1}{@{}c@{}}{\textbf{S}}  & \textbf{V} & \textbf{P S}& &&  \\
\hline 
\citet{52} & Transformer & \Checkmark & \Checkmark & \XSolidBrush & \XSolidBrush &Twitter15 Twitter16 &Transformer captures the reliance between tweets and retweets &Breaking the propagation tree structure, the retweet information in different conversation threads can be calculated for correlation\\

\citet{82} &Transformer &\Checkmark & \Checkmark & \XSolidBrush & \XSolidBrush & SemEval2017 PHEME &Transformer extracts position feature&Using tags for position detection makes the results of rumor detection more accurate\\

\citet{83} &Transformer &\Checkmark & \XSolidBrush & \XSolidBrush & \XSolidBrush &PolitiFact GossipCop PHEME &Transformer's coding layer term news coding &External knowledge using knowledge graph \\

\citet{86} &Transformer &\Checkmark & \XSolidBrush & \Checkmark & \XSolidBrush &CLEF-EN CLEF-Ar MediaEval LESA&Transformer extracts text features&Explore the importance of images in rumor detection\\

\citet{87} &Transformer &\Checkmark & \Checkmark & \XSolidBrush & \XSolidBrush & Twitter PHEME &Transformer extract position and attitude feature and communication feature&Tree Transformer to preserve the user interaction of the dialog\\

\hline
\end{tabular}
\begin{tablenotes}
\item DL:Deep Learning; T:Text; S:Social; V:Vision; P S:Propagation Structure
\end{tablenotes}
\end{table}
\end{center}

{\subsection{Rumor Detection Based on Other Structures}}

{There are much other deep learning structures\citep{other1,other2}. In addition to the above methods, there are studies to build their algorithms through other model structures\citep{93,79} and obtained good experimental results. For example, in the work of \citet{93}, XLNet is used to map the contextual content of rumors to a high-dimensional to learn the content features of rumors and combined with the rumor topic distribution of LDA(Latent Dirichlet Allocation) to solve the problem of rumor detection for COVID-19. Since this method focuses too much on the content itself, it cannot guarantee the outbreak of new rumors in the new crown era. However, in order to solve the problem of difficulty in the communication of rumor features between different domains, \citet{79} separated the domain-specific features and non-domain-specific features in the rumor text content through a fully-connected network. Weakly supervised classification on unlabeled datasets of range domains. Nevertheless, they ignored the full convergence speed and high computational complexity of the fully connected network.}

\section{Rumor Detection Methodology}

{To answer RQ3}, in this chapter, we classify rumors detection methods based on the newest work and introduce various novel methods used in the newest research to detect rumors. After experiments, these methods are better, more efficient, and more automated than traditional manual features methods. The proportion is higher. The structure of this article is shown in the figure 5.And summary about the deep learning architectures, tools/libraries and performance matrices used in research studies in Table 6.

\begin{figure}[!h]
\centerline{\includegraphics[scale=0.52]{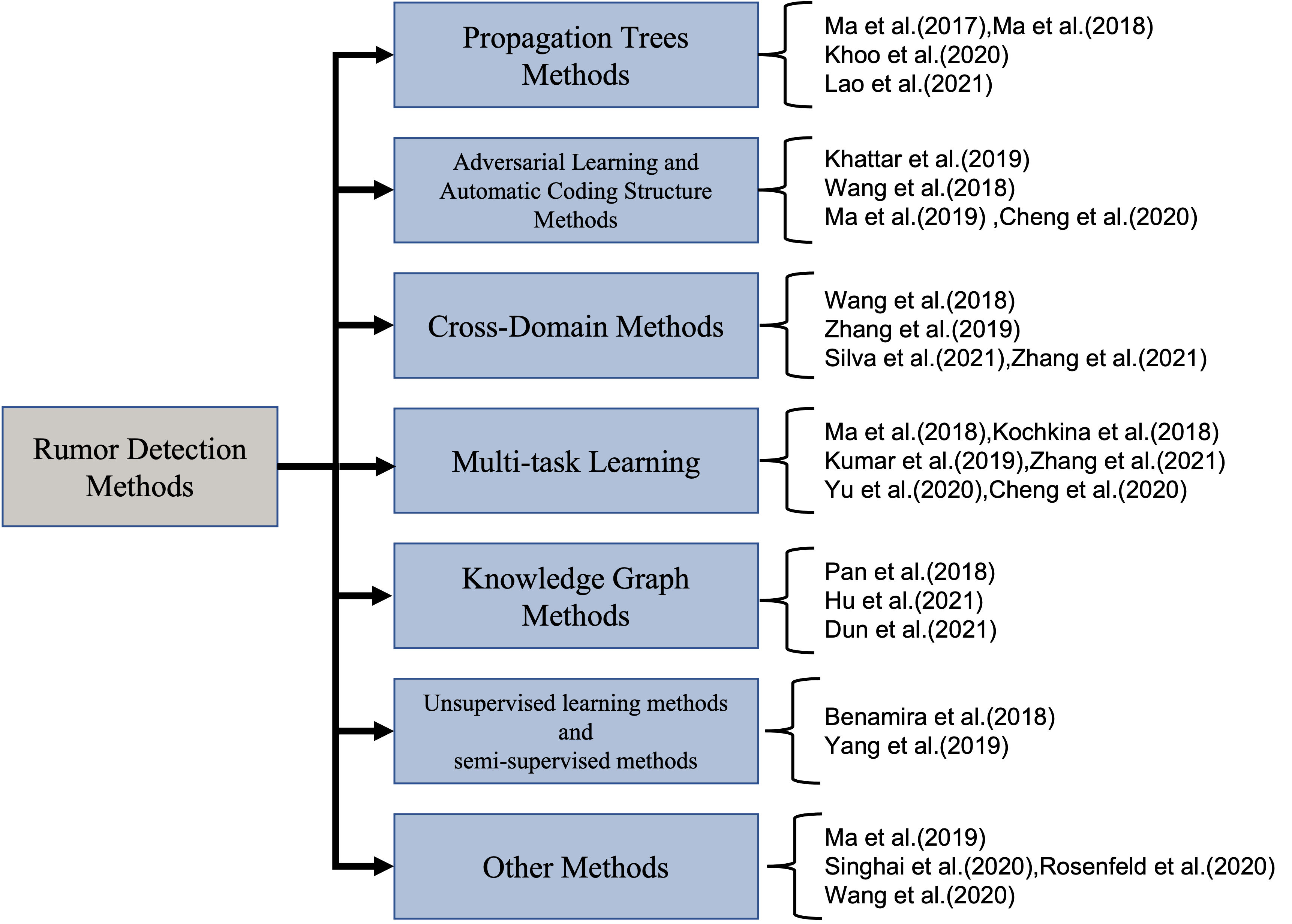}}
\caption{ {Classification and related work of the latest methods of rumor detection}\label{fig4}}
\end{figure}

\begin{sidewaystable}
\begin{center}
\centering
\tiny

\begin{longtable}{p{2.5cm}p{1cm}p{5cm}p{1cm}p{7cm}}

\multicolumn{5}{l}{\small{\textbf{{Table 6}}}}\\
\multicolumn{5}{l}{\small{{Detail of deep learning research studies in section 5.}}}\\

\hline
\label{table}\\

\textbf{Study}&\textbf{DL Method}&\textbf{Performance metrics}&\textbf{Tool/ Library} & \textbf{Desctiption of Architecture} \\
\hline 
{~\citet{21}} & {RvNN }& {Acc.:0.723 NR:F1:0.682 FR:F1:0.753 TR:F1:0.821 UR:F1:0.654 }&{ theano} & {They input each propagation node into the RNN and propagate the features along the path of the propagation tree to the next RNN Unit.}\\
{~\citet{68}} &{ RNN GNN} & {Acc.:0.925 Pre.:0.9245 Recall:0.925 F1:0.9247} & {-} & {Graph convolutional neural(GCN) network to extract non-Linear structure features ,LSTM extract linear sequence feature.}\\

{~\citet{53} }&{RNN Transformer }& {Acc.:0.85 NR:Acc:0.844 FR:Acc.:0.857 TR:Acc.:0.883 UR:Acc.:0.814} & {pytorch tqdm sklearn}  &{Use RNN to learn context and use Transformer to find tightly related conversation threads}\\

~\citet{35} &{ RNN }&{Acc.:0.781    \textbf{Rumor:} \qquad  Pre.:0.796  \qquad Rec.:0.796 \qquad F1:0.784  \qquad  \textbf{None-rumor: } \qquad Pre:0.791 Rec.:0.766 F1:0.778 }&{theano} & {Gated Recurrent Unit used for representing hidden units}\\

~\citet{24} & {CNN RNN} &{Acc.:0.827 Pre.:0.847 Recall:0.812 F1:0.829} & {sklearn pytorch scipy  \qquad jieba Word2Vec} &{TextCNN extracts text features and VGG extracts visual feature }\\

~\citet{43} & {RNN CNN } &{ Acc.:0.824 \textbf{Fake:} Pre.:0.854 Recall:0.769 F1:0.809  \qquad  \textbf{Real:} Pre.:0.802 Recall:0.875 F1:0837 } &{ keras sklearn Word2Vec nltk } &{Bi-LSTM as both encoder and decoder to extracts text features and VGG-19 with double fully connect layers extracts images features.}\\

~\citet{55} &{ RNN }& {Acc.:0.752 Pre.:0.755 Recall:0.752 F1:0.752} &{ - }& {Bidirectional LSTM Unit as the hidden units of encoder-decoder.And learn multi-task features by Bi-LSTM.}\\

~\citet{79} &{ Dense }&{Acc.:0.877 Pre.:0.840 Recall:0.832 F1:0.836 }&{tensorflow keras sklearn nltk }&{ Dense Net as the domain-specific discriminator,the domain-shared discriminator and generator.}\\

~\citet{44} &{ RNN CNN }& {Acc:0.763 }&{ - }& {Use GRU as the hidden unit of decoder-encoder; Residual Blocks as content encoder and decoder to Extract visual features.}\\
 
~\citet{45} &{ RNN CNN }& {Acc:0.866 \textbf{Rumor}: Pre.:0.893 Recall:0.847 F1:0.870 \textbf{Non-Rumor:} Pre.:0.866 Recall:0.867 F1:0.866 }& {-} & {VGG-19 extract image feature and Bi-GRU extract text feature. The rumor characteristics between different posts are transmitted through GRU,and Separating domain specific features with event memory network.}\\

~\citet{55} & {RNN} & {F1:0.405 Acc.:0.405} & {keras} & {LSTM is used to learn text features, features of rumor detection task, stance detection features and rumor verification features.} \\

~\citet{74} & {RNN} & {Acc.:0.638 F1:0.606 }& { - } & {LSTM unit is used to Rumor Verification and Stance detection.}\\

~\citet{85} &{ RNN }& {F1:0.464 NR:F1:0.876 FR:F1:0.543 TR:F1:0.114 UR:F1:0.333 }& {theano} & {Use GRU to merge the characteristics of the two tasks.}  \\

~\citet{86} &{RNN BERT CNN }&{Acc.:0.819 Pre.:0.75 Recall:0.8667 F1:0.8041} & {pretrained BistilBERT } & {word embedding and image embedding by pretrained BistilBERT and VGG19. Use GRU as the hidden layer unit in Meta Multi-task layer.}\\

~\citet{83} &  {Transformer} & {Pre.:0.8687 Recall:0.8499 F1:0.8539 Acc.:0.8586 AUC:0.9197 }& {Wikipedia  TagMe Word2Vec} & {Transformer as encoder to encode entities, news, and entity contexts sequence.}\\

~\citet{hu2021compare} & {GNN RNN }& {F1:0.8912 Prec:0.8982 Recall:0.8917 }& {pytorch Word2Vec functools  TagME }& {GCN learns the structural characteristics of heterogeneous graphs with attention mechanism and use LSTM to embed information describing entities.}\\

~\citet{93} & {CNN XLNet} & {Acc.:0.846} &{keras pandas sklearn} & {Use Pretrained XLNeT to embed the title and content to get the text features.Use VGG-19 to embed the image to get visual feature.}\\

~\citet{94} & {CNN} & {Acc.:0.824 AUC:0.873} &{ pytorch }&{TextCNN to get text embedding.}\\

~\citet{95} &{ RNN }& {F1:0.807 True:Pre.:0.637 Recall:0.665 F1:0.651 False:Pre.:0.874 Recall:0.860 F1:0.867 }& {theano} &{GRU with attention mechanism for claim verification.}\\

\hline

\end{longtable}

\begin{tablenotes}
\item{DL:Deep Learning; NR: nonrumor; FR: false rumor; TR: true rumor; UR: unverified rumor; Pre.:Precision; Acc.:Accuracy.}
\end{tablenotes}
\end{center}
\end{sidewaystable}

\subsection{Propagation Trees Method}

\begin{figure}[!h]
\centerline{\includegraphics[scale=0.5]{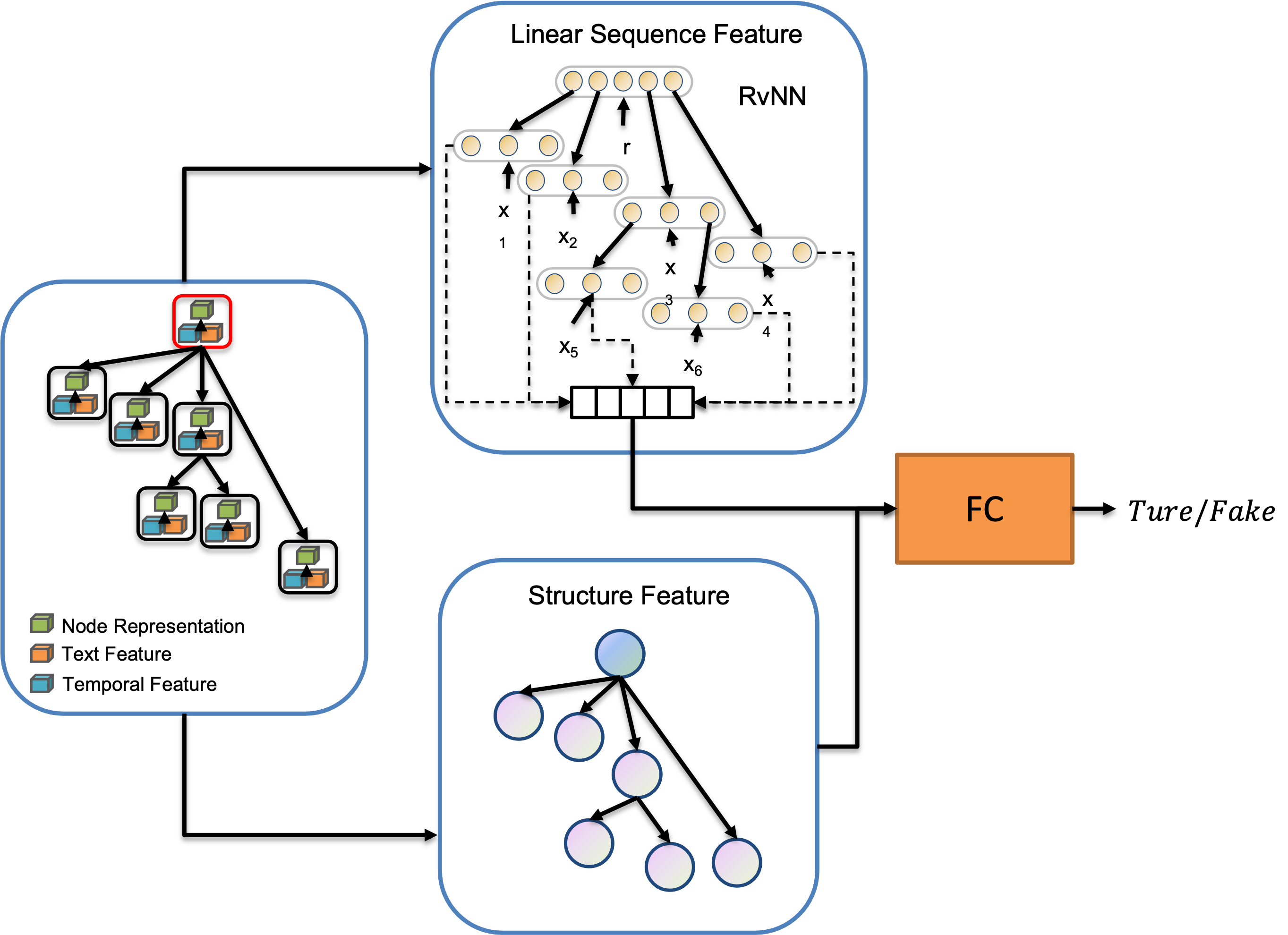}}
\caption{{Architecture of Propagation Trees Method.Top-down RvNN is used as an example to extract the linear sequence features of Twitter propagation. Some studies have taken the nonlinear structural features of the propagation tree and fused with the linear features to classify rumors.}\label{fig6}}
\end{figure}

Since tweets in social media will lead to many comments, and users can further comment on the comment information, thereby forming a variety of branches. In this way, ~\citet{60} modeled the spread of rumors in the form of a tree, as Fig6. ~\citet{22} identified rumors by capturing the substructure of the propagation tree to evaluate the similarity between the propagation trees. However, this study only compares the structural similarity between propagation trees. Afterwards, ~\citet{21} improved ~\citet{22} and established a tree-based recurrent neural network model, taking a propagation tree rooted at the source post as input and passing a bottom-up and top-down tree structure recurrent neural network {(Figure 6 uses a top-down RvNN as an example) }jointly captures the linear sequence feature . Compared to ~\citet{22}, this method does not require comparison the structure of propagation tree.

{~\citet{68} mentioned a combination of nonlinear structure learning and linear sequence learning to learn the propagation characteristics in the propagation tree. This study uses nonlinear structure learning to learn diffusion propagation features along the diffusion path of the propagation tree and uses the linear sequence to learn to aggregate the features of context nodes and retains temporal features to represent sequential propagation. The graph convolutional neural network is used to extract the nonlinear propagation structure features, and LSTM is used as the linear feature learning that retains the timing. Finally, combining two features to classify rumors and explore the relevance.}

{However, different from the methods mentioned above,~\citet{53} flattened the structure of the propagation tree, allowing all possible pairwise interactions between tweets, and used the self-attention mechanism to obtain two significantly related tweets from a large number of tweets. Because they believe that although the tree model can model the structural information that exists in the dialog thread, and the information is passed from parent to child in the tree model, each user can usually observe all the replies in different branches of the dialog. The content that debunks fake news may be tweets used in other tree branches. However, although their approach can link the branch information of different propagation trees, it destroys the structure of the propagation tree and cannot use structural features.}

{Although they (\citet{22},\citet{21},~\citet{68},~\citet{53}) use structure and content information, not every propagation path is essential. Some users are easily affected by rumors, and their comments may affect the judgment of rumors. Therefore, they lack information on users. The use of attributes to assign the weights of propagation paths to different users.In addition, there is a lack of work on the changing trend of public opinion in the use of the communication path. Because many comments believe that rumors will affect subsequent comment users, it will cause the user's position in the communication path to change, which is also the trend of public opinion.}

\subsection{Adversarial Learning and Automatic Coding Structure Method}

\begin{figure}[!h]
\centerline{\includegraphics[scale=0.5]{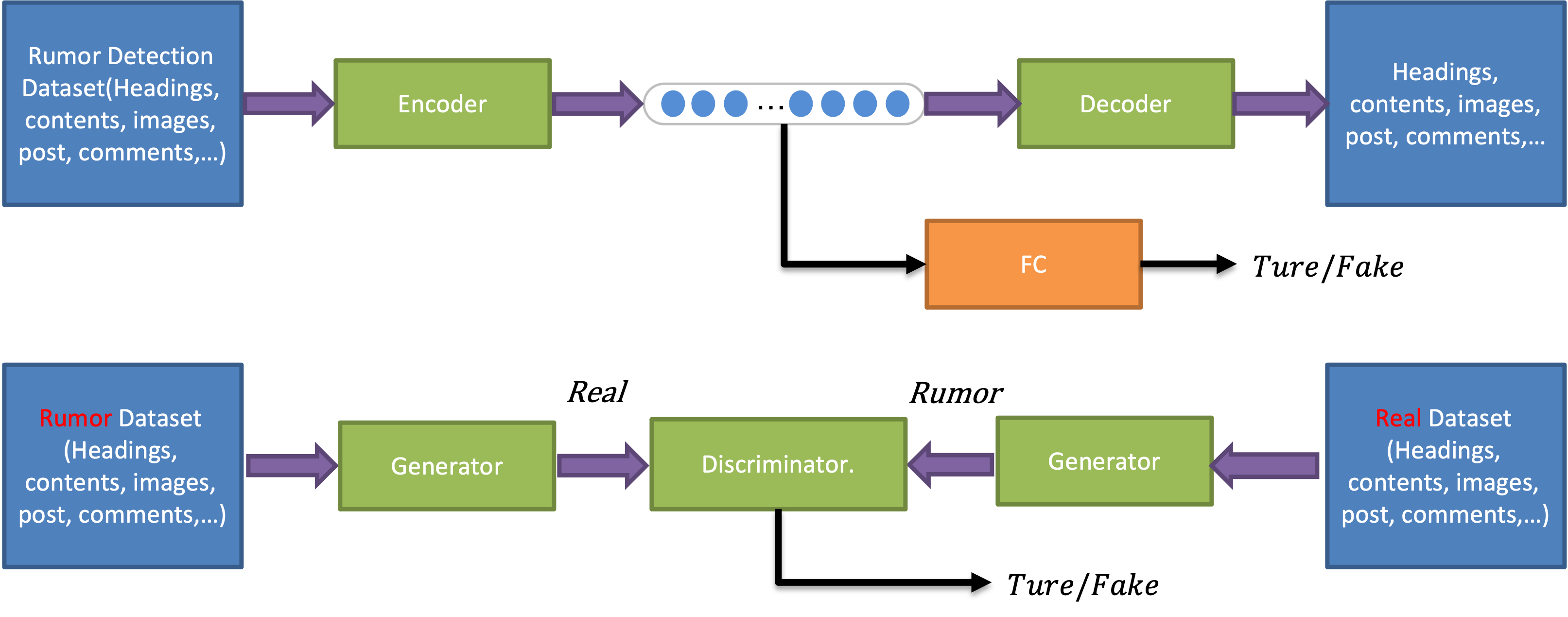}}
\caption{{Architecture of Adversarial Learning(below) and Automatic Coding Structure(above) Method}\label{fig7}}
\end{figure}

{Figure 7 shows the rumor detection framework of Adversarial Learning and the rumor detection framework of Automatic Coding Structure. Among them,  Adversarial learning generally uses generators to generate real information and rumors from rumors and real information to train the discriminator to judge the features of rumors in all directions. After the final training, the discriminator is used to distinguish between rumors and real information. The rumor detection method of the self-encoder usually uses an encoder to encode the rumor data set into a feature vector, and tries to restore the feature vector to the original data through the decoder structure opposite to the previous encoder, and feeds back the generated data to the encoder by comparing the generated data with the original data. Finally, the feature vector generated by the well-trained encoder is classified into rumors through the fully connected layer and activation function.}

~\citet{35} pointed out that rumor makers expand the spread of rumors, which poses greater challenges to such data-driven methods. Therefore, using the characteristics of generative adversarial networks, generators are used to simulate rumors and propaganda to output more challenging examples, so as to promote the discriminator to strengthen the feature learning of such difficult examples to capture more discriminative patterns. The research uses GRU as generator to get representation of rumors. In the model, the generator is encouraged to generate motion-like instances to fool the discriminator, and the discriminator focuses on learning more discriminative features. The discriminator can master more discriminative features by using the data features combined by the generator, especially the information of non-trivial patterns learned from low-frequency patterns. {Later,~\citet{24} proposed a cross-domain rumor detection framework, EANN{(Event Adversarial Neural Networks)}, in which a multi-modal feature extractor was used as a generator to generate rumor feature vectors to deceive the event discriminator in the framework. Expect to abandon the characteristics of events in a particular field. On the other hand, the event discriminator tries to find the information related to the specific event contained in the feature representation to identify the event.}

~\citet{43} got inspiration from EANN and built an end-to-end network with variational auto-encoder called Multimodal Variational Autoencoder (MVAE), using a dual-modal autoencoder and a classifier to complete Rumor detection task. It mainly includes three parts: encoder, decoder and detector. Encoders and decoders use the idea of adversarial learning. The encoder learns features from both text and image, and then encodes them as vectors. The decoder takes the output vector of the encoder as input and decodes it into text and image features. And through the decoded features and the input features of the encoder to calculate the loss to train the encoder and the decoder, and finally complete the rumor detection judgment through the detector.~\citet{55} took inspiration from MVAE and created a text-based multi-task auxiliary variational autoencoder. In this work, a LSTM-based variable autoencoder model was proposed to pass the text The RNN encoder with LSTM unit performs encoding, and then uses the same parameter RNN network for decoding.

{The current self-encoding method or adversarial learning method for rumor detection only rests on restoring the image and text features but lacks the generation of structured data sets, such as propagation trees and propagation graphs. Although \citet{yang2021rumor} has used graph structure adversarial learning for the learning task of propagating graphs, they only consider the detection of abnormal propagation points to help the rumor detection task.}

~\subsection{Cross-Domain Method}

\begin{figure}[!h]
\centerline{\includegraphics[scale=0.5]{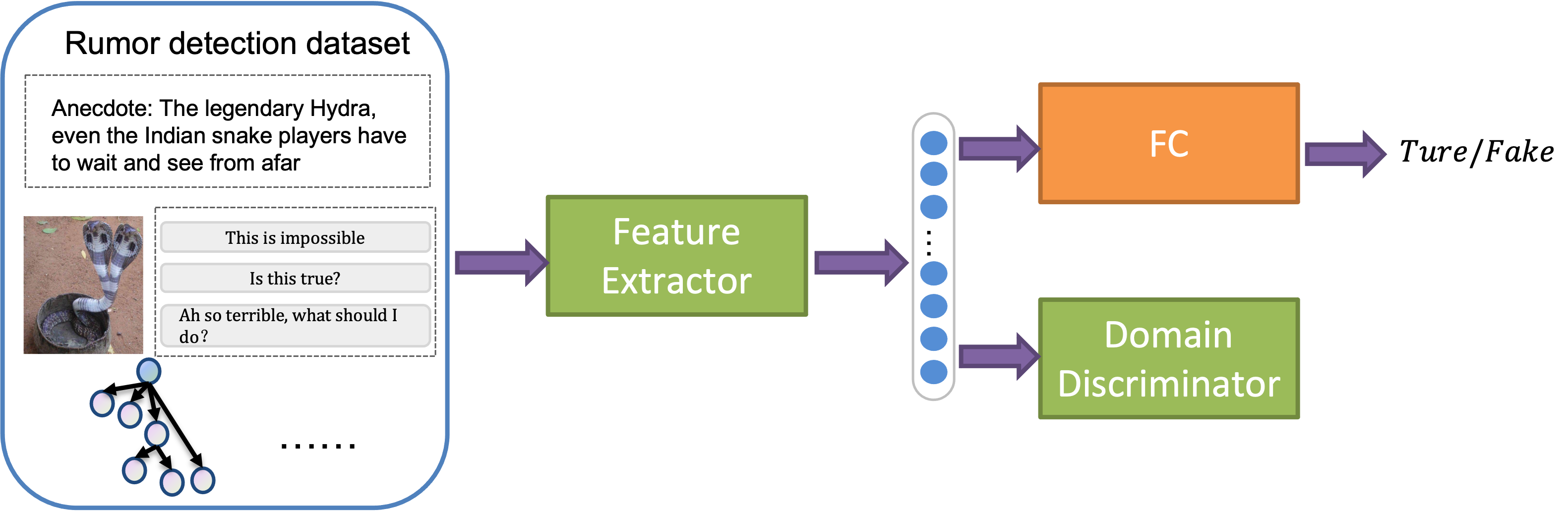}}
\caption{{To improve the performance of cross-domain rumor detection, the primary research is to remove the domain-specific features in the rumor feature. After the features are extracted, the domain discriminator is used to improve the performance of the feature extractor to remove the features of the specific domain. Regardless of how the latest research assists in domain elimination tasks, the primary method used remains in adversarial learning.}\label{fig8}}
\end{figure}

Rumor detection methods were based on specific events or specific areas of learning, resulting in failure to identify fake news in real-world news streams. For example, specific nouns in the fields of politics, entertainment, medical care, etc. will cause the training results to perform poorly in other fields. However, research hopes that a rumor detection method can be truly applied in the real world. ~\citet{84} also showed in their research that most fake news detection techniques are not good at identifying fake news from rare fields and recording their characteristics during training. Therefore, there are many studies aimed at solving how to retain specific domain and cross-domain knowledge in news and rumor records to detect fake news in cross-domain news data sets.

~\citet{24} pointed out that most early models have poor monitoring performance for emergencies and cannot capture the feature of different events that are not shared in specific events. To solve this problem, we propose to extract cross-domain features from multi-modal data of fake news. Use the adversarial learning method mentioned to remove the specific event features in the multi-modality and keep the feature representation of the event unchanged. The effectiveness of cross-domain rumor detection has been improved.

~\citet{79} proposed a framework to save news features in specific fields jointly and cross-fields to detect false news from different fields. The framework integrates an unsupervised domain embedding learning module and a regulated and domain-independent news classification module. The unsupervised domain embedding learning module uses multi-modal tasks to express news as low-dimensional representation and give labels to unlabeled news through learning. The classification module learn specific domain and cross-domain knowledge in news while identifying fake news. The research maps the representation of news to two parts for learning specific domain knowledge and cross-domain knowledge respectively.

In 2021, ~\citet{44} were inspired by ~\citet{24} to conduct cross-domain event fake news detection work for multi-modal data and introduced the multi-modal separation representation learning in study of ~\citet{24} to social media rumors detection to explore the commonalities and characteristics of different modalities across domains. A multi-modal unsupervised domain-adaptive method is proposed, which can derive event-invariant features, conducive to detecting rumors about emerging social media events. In the research, the rumor style feature classifier and content style classifier are trained on text and images respectively to extract rumor style and content features from visual information and content information. The unsupervised domain adaptation module based on adversarial learning learns the characteristics of the transferable rumor style in multimedia posts and transfers the knowledge obtained from historical events to new events.

{Unlike the above work, ~\citet{45} extracted entity knowledge from the outside and record it in shared storage space. When a high-level representation of an emerging event is given, the event memory network will retrieve it from the external knowledge base and output the features of entities existing in events in the same field or features of rumors with similar entities.}

{The work mentioned above is from removing specific domain features or extracting external knowledge to retain the general domain features of rumors as cross-domain rumor detection methods. Although these methods solve the interference problem of domain nouns in some memorable domains, in the research process Ignoring the different ways of dissemination of rumors in different fields because many users have limited understanding of specific fields, the spread of rumors has accelerated, and users who question rumors are even rarer.}

\subsection{Multi-task Learning}

\begin{figure}[!h]
\centerline{\includegraphics[scale=0.5]{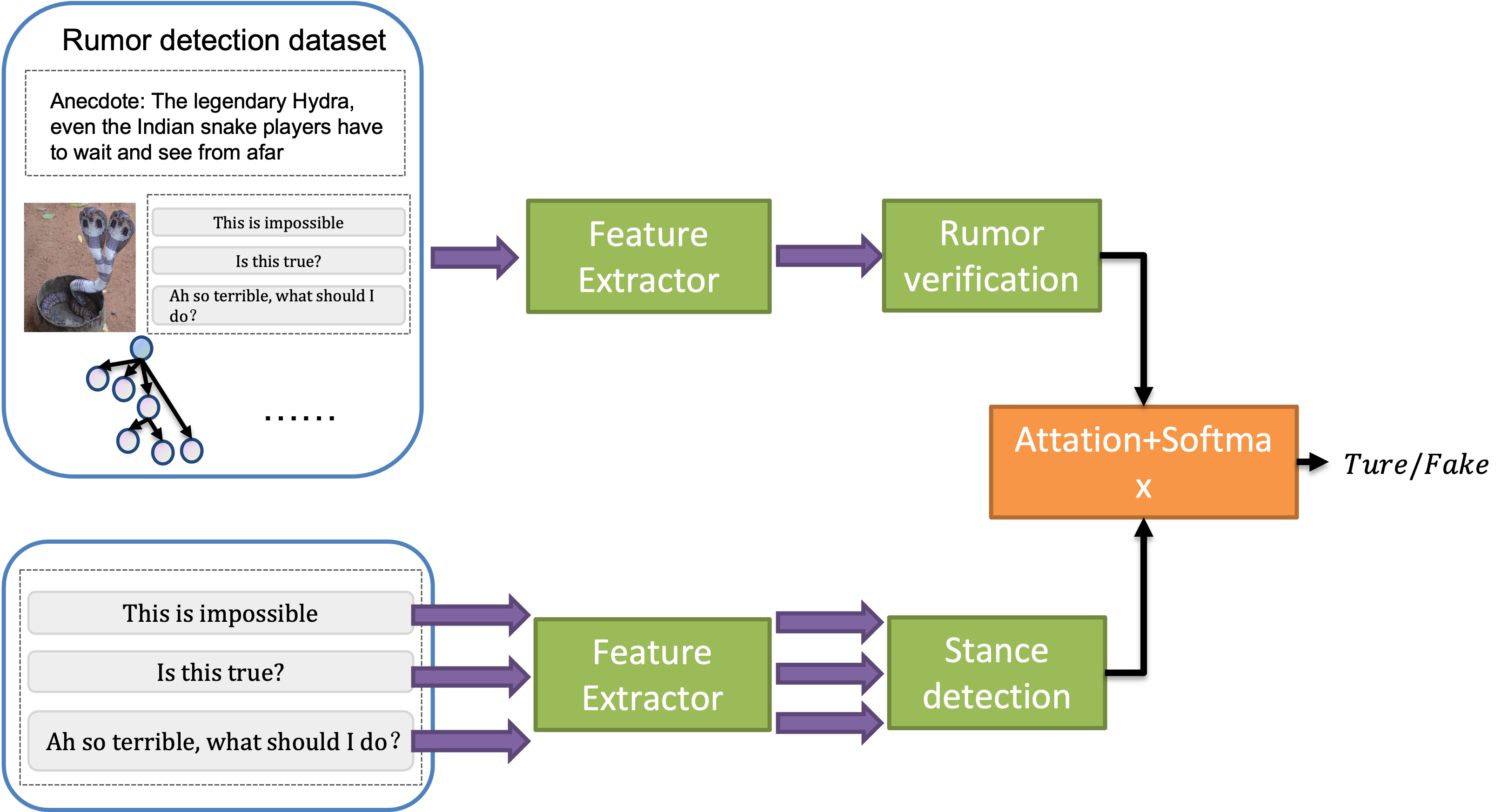}}
\caption{{Architecture of Multi-task learning for Rumor Detection. Including two tasks: Stance Detection and Rumor Verification. These two tasks are also the choice of most multi-task rumor detection. Helping rumor detection tasks by detecting the stance of comments.}\label{fig8}}
\end{figure}

Multi-task Learning is a machine learning method opposite to Single-task Learning. In traditional machine learning, the standard algorithm theory is to learn one task when the output of the system is a real number. Multi-task learning is a kind of joint learning, where multiple meta-tasks are learned in parallel so that the learning results influence each other. That means multi-task learning is to solve multiple meta-problems at the same time. In the latest research, studies are using multi-task learning methods to solve rumor detection (~\citet{23,55,74,85,86}).

In the latest work, a large amount of literature uses user comments and evidence to extract the user's position on rumors. This type of work is called position detection. Joint stance detection task and rumor classification task, and then share the features extracted from the two tasks for rumor detection, {as Fig.9.} Many experiments have proved that position detection has a huge positive impact on rumor detection.~\citet{85} and ~\citet{86} used GRU to optimize stance detection jointly, and rumor classification tasks in their work, and both achieved significant improvements. {However, they assign the same weight to each user’s comment, ignoring that some untrusted users’ comment characteristics have little or even a negative effect on rumor detection.}.Since not all users have high credit, so ~\citet{74} added user confidence features and text feature embeddings in their work. {Nevertheless, it still ignores the structure of the comment information} ~\citet{77} proposed a tree-structured LSTM multi-task model with convolutional units and used the tree structure to propagate effective position signals upwards to classify rumors at the root node. However, the above research did not correlate the results calculated by position detection with the specific output layer of rumor detection, so  ~\citet{82} added a Transformer layer to the multi-task framework of joint position detection and rumor classification to capture one of the blog posts in the entire conversation thread. By combining two transformer components to solve the problem of not explicitly modeling the interaction between the position-specific layer and the rumor-specific layer in previous work.

{For the multi-task learning method that uses stance detection to help rumor detection, although users’ feedback on rumors can be used to make preliminary judgments on rumors, we believe that this method negates the role of public opinion trends. The dissemination harm of rumors is the pressure of public opinion trends. Some users who are not firm will shake their stances when observing the positions of the public. Stance detection will have a counterproductive effect on this phenomenon, which will mislead the task of rumor detection. Although some studies have used user credibility information, they still have not made any countermeasures against changes in public opinion trends. Therefore, this is also a challenge to the multi-task learning method using stance detection.}

{The framework of ~\citet{55} is different from that shown in Figure 9. The reason is that they were inspired by ~\citet{10} that they classified rumors into four parts: rumor detection, rumors Tracking, position classification, and accuracy classification. }
So, ~\citet{55} took these four parts as four meta-tasks and established Rumor Detector, Rumor Tracker, Stance Classifier, and Veracity Classifier to target the above four tasks respectively and combine the results of the four classifiers to assist the rumor detection task.

\subsection{Knowledge Graph Method}

\begin{figure}[!h]
\centerline{\includegraphics[scale=0.5]{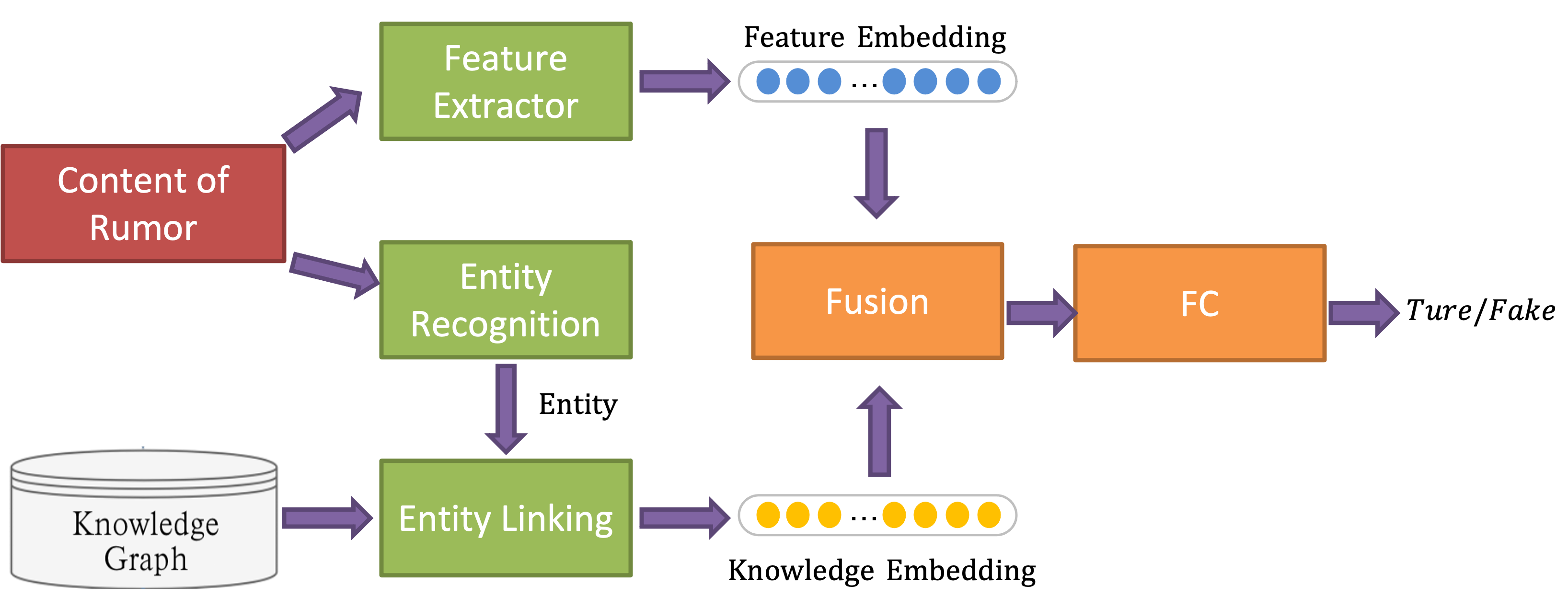}}
\caption{{Architecture of Knowledge Graph(KG) Methods for Rumor Detection. The steps of current general method for rumor detection based on knowledge graph : first identify the entity of the content of the rumor, then entity link to obtain external knowledge from knowledge graph , then represent the knowledge to vector, last classify the rumors after fusion or comparison with the original features.}\label{fig10}}
\end{figure}

The knowledge graph is a semantic network that reveals the relationships between entities. There are millions of items describing real-world entities, such as people, places, and organizations. In the knowledge graph, entities are represented as nodes, and the relationships between these nodes are described as edges. Knowledge graphs are widely used through entity links, such as movie recommendation (~\citet{88}), machine reading (~\citet{89}), text classification (~\citet{90}), etc. In the latest work, some documents extract the entity nouns in the rumors and then obtain knowledge other than the rumors through the knowledge graph to assist the rumor detection task.

~\citet{91} first proposed a content-based fake news detection method using knowledge graphs and created three types of knowledge graphs: based on fake news databases, open knowledge graphs, and reliable news organizations. The actual news library. And use the triple information extracted from the fake news to retrieve the created knowledge graph, represent the triple in the vector space, and judge the truth of the news article through the vector.{However, this method can only record news events that have occurred, and its ability to detect new rumors is minimal.}

~\citet{83} integrated the external knowledge language news detection from the knowledge graph, identified the entities in the news content, and matched them with the entities in the knowledge graph. The context in the knowledge graph is used as external knowledge for supplementary information, and each entity is given a weight to express its importance.{However, they embed the context in the knowledge graph together. Since new knowledge may be missing in the knowledge graph in new outbreaks, the impact on the rumor detection task is not stable.}

{~\citet{hu2021compare} proposed a CompareNet, which uses the knowledge in the news knowledge graph to align with the entities of the news text to find the corresponding entity in the knowledge graph, that is, context information. The structural information and text knowledge of the joint knowledge map is compared with the description of the entity in the original news, and similar features are calculated, and the high-dimensional features of the original news are combined to classify news. Compared with \citet{83} ,their method compares the embedded information and structural features of the original text to get better results in entity comparison. However, they lack the consideration of multiple feature fusion weights in the feature fusion process.}

\subsection{Unsupervised and Semi-supervised Learning  Methods}

Unsupervised learning and semi-supervised learning are two typical methods in machine learning. The unsupervised method is due to a lack of sufficient prior knowledge and has to solve the problem in pattern recognition based on the training samples of the location category. Semi-supervised learning is a learning method that combines supervised learning and unsupervised learning. Semi-supervised learning uses a large amount of unlabeled data while using labeled data for pattern recognition. Because the relevant data sets for rumor detection are less labeled and involve a wide range of fields, the professional requirements of the labeled personnel are higher, making it challenging to label data sets for rumor detection. Therefore, there are works for unsupervised learning methods and semi-supervised learning methods for rumor detection.

Because there are fewer tagged articles in the current popular data set, and the scope of inclusion is relatively narrow. However, there is the possibility of inaccuracy in marking news in a crowdsourced manner. ~\citet{25} mapped the text to the latent representation in the Euclidean space and proposed a semi-supervised learning method combining the K-nearest neighbor algorithm with the graph neural network and the graph attention neural network to classify the text into rumors and non-rumors.

~\citet{27} proposed an unsupervised generation method for social media fake news detection, using auxiliary information of users participating in news tweets on social media, extracting users’ opinions on news, and collect opinions in an unsupervised way and generate estimates result. The authenticity of news and the credibility of users are regarded as potential random variables, and users’ participation in social media is used to determine their views on the authenticity of the news. And use the Bayesian network model to capture the conditional dependence between news authenticity, user opinions, and user credibility.

{Since rumors in social media are designed in many fields and require a large number of domain experts to classify rumors, the data set labeling of rumors detection is a very difficult and requires a lot of money. In theory, unsupervised learning methods And semi-supervised learning methods can solve this kind of problems very well(\citet{25},\citet{27},\citet{jahanbakhsh2021semi}). However, although there have been some studies, there are still no outstanding applications.}

\subsection{Other Methods}
In addition to the rumor monitoring methods mentioned in 5.1-5.6, there is also much work using new technologies for rumor monitoring tasks. For example: transfer learning (~\citet{93}), reinforcement learning (~\citet{94}), etc.

~\citet{93} uses a multi-modal method that uses transfer learning to obtain semantic and contextual information from news articles and related images to improve fake news detection accuracy. The research uses pre-trained language embedding and image network models to extract features. These feature vectors are input into fully connected layers for classification.

~\citet{94}  proposed a weakly-supervised framework based on reinforcement learning. The framework can use user reports as weak supervision to expand the training data used for fake news detection. The framework contains three components: an annotator, a fake news detector, and an enhanced selector. Using a set of labeled fake news samples and user feedback on these news articles, the framework can train annotators based on the feedback and automatically assign weak tags for user feedback to unlabeled news articles based on the content. Using reinforcement learning technology, the reinforcement selector selects high-quality samples from the weakly labeled samples as the input to the fake news detector. The fake news detector finally assigns a label to each input article based on its content.

~\citet{96} studied the effect of separate diffusion mode on rumor detection, used graph kernel to extract complex topological information from Twitter cascade structure, and trained a predictive model that ignores language, user identity and time, and proved the removal for the first time. The diffusion model of complex information has high accuracy. The results show that through proper gathering, even in the early stages of dissemination, the spread pattern of rumors in the crowd may reveal a powerful signal of the truth and falsehood of the rumors, and at the same time prove that the spread pattern can indeed provide the accuracy of the rumors.

~\citet{95} proposed a sentence-level embedding method for rumor detection based on hierarchical attention networks, which integrates the implicated relationship between each sentence of the rumor and the evidence and ensures the consistency of the evidence. The end-to-end hierarchical attention network designed in the work The end-to-end hierarchical attention network is used for sentence-level evidence embedding. The purpose is to focus on important sentences/evidence by considering the topic coherence and semantic reasoning strength. The model can more reasonably determine the rumor’s verdict and embed the evidence sentence into the learned statement. At the same time, with the help of attention, key evidence can be highlighted and cited to explain the judgment better.

\section{Dataset}
{To answer the RQ4, the types of datasets available for rumor detection research are introduced in detail with this section. } The detailed information of the data set is summarized and plotted in Table 7, in which only the relevant data set information given by the original provider of the data set is provided. If the number is deleted and modified by other work, no statistics will be made.

Most of the work uses data sets from Twitter, Weibo, and news in the latest research. Some of these data sets were obtained by researchers based on the public APIs of social media. For example, ~\citet{22} used their published APIs to collect data from Twitter to establish the Twitter15 and Twitter16 data sets, which contain 1381 and 1181 propagation trees, respectively. And use four labels to annotate each tree: non-rumors, false rumors, true rumors and unverified rumors. 

In addition, jobs are using publicly available news event data sets. For example: MediaEval(~\citet{98}), Buzzfeed Election(~\citet{99}), PHEME(~\citet{100}), LIAR(~\citet{34}), etc. Among them, PHEME uses manually marked rumors and non-rumors, including original blog posts and replies about nine breaking news events. The Buzzfeed Election dataset includes complete Facebook news from September 19 to September 27 in the 2016 U.S. election. LIAR contains fake news from the PolitiFact collection, including short statements, in the form of press releases, TV news, etc.

However, in the latest research, researchers have found that most of the existing data sets are small in scale or contain only a few areas of fake news. For this reason, many works have proposed new large-scale, multi-domain data sets. For example, recent work proposes a large-scale, multi-modal data set NewsBag(~\citet{101}), containing 200,000 real news and 15,000 fake news. The real news comes from The Wall Street Journal, and the fake news comes from The Onion. ~\citet{104} believe that some fake news is maliciously modified or distorted in the process of multiple dissemination of real news. If this situation is ignored, the model will be deceived. Therefore, a new data set to track the evolution of fake news is released, called the Fake News Evolution (FNE) data set. The data set comprises 950 paired data, and each data set is composed of articles representing three important stages of the evolution process, namely, truth, fake news, and evolved fake news.

There are also studies focusing on claim extraction and verification. For example, ~\citet{103} disclosed a data set called FEVER, which consists of 185,445 statements generated by modifying sentences extracted from Wikipedia, and then without knowing the source of these sentences authenticating. These statements were classified as supporting, refuting, and insufficient information by the commenter. And provide the annotator's judgment basis in the data set to judge the first two types of information.

\begin{sidewaystable}

\begin{center}
\centering
\tiny

\begin{longtable}{p{1.8cm}p{1.2cm}p{1.8cm}p{3cm}p{1.2cm}p{0.8cm}p{7cm}}

\multicolumn{7}{l}{\small{\textbf{Table 7}}}\\
\multicolumn{7}{l}{\small{Detail of dataset.}}\\

\hline
\label{table}\\

\textbf{Dataset}&\textbf{Reference}&\textbf{Size}&\textbf{Label} & \textbf{Number for labels} & \textbf{Data type} & \textbf{Detail of dataset}\\
\hline 

PHEME& \citet{100} & 5802 tweets &rumor /none-rumor&1792/3830&text&{It contains five events: Charlie Hebdo, Ferguson, Germanwings Crash, Ottawa Shooting, Sydney Siege. It's \textbf{imbalanced dataset.}}\\

Twitter15& \citet{22}&{1380 tweet-tree} &non-rumor /false-rumor /true rumor /unverified&374/370/ 372/374 &Tree-structure text & {It contains 276663 users, 1490tweets, 331612 threads. It's \textbf{balance dataset}}.\\

Twitter16& \citet{22}&{1181 tweet-tree} &non-rumor /false-rumor /true rumor /unverified &205/205/ 205/203 &Tree-structure text &{It contains 173486 users, 818 tweets, 204820 threads. It's \textbf{balance dataset}}.\\

MutiSource-Fake &  \citet{32} &{11397 news }&fake/real &5403/5994 &text &{ It contains news from OpenSources.co, MediaBiasFactCheck.com, PolitiFact news websites’ lists. It's \textbf{balance dataset}}.\\

Weiibo &  \citet{19} & 146 events, 50287 tweets  & rumor /none-rumor &23456/26257 &text, image, users &{Real world multimedia dataset from Sina Weibo contains 50287 tweets and 25953 images and 42310 users. It's \textbf{balance dataset}}.\\

MediaEval& \citet{98}&{15821 tweets}& rumor /none-rumor &9596/6225 &text, image &{ It contains tweets related to the 11 events, comprising in total 193 cases of real and 220 cases of misused images and videos, associated with 6225 real and 9596 fake tweets posted by 5,895 and 9,216 unique users respectively. It's \textbf{imbalanced dataset}.}\\

FakeNewNet&  \citet{shu2020fakenewsnet} &23196 news &fake/real &5755/17441 &{text, image, users, network, response }&{This dataset contains 23196 news, 19200 images from PolitiFact and GossiCop. It's \textbf{imbalanced dataset.}} \\

Buzzfeed Election & \citet{99} &71 news &fake/real &35/36 &text &{It collected the news stories found in Buzzfeed’s 2016 article on fake election news on Facebook (Silverman 2016). It's \textbf{balance dataset }}\\


LIAR &  {\citet{34}} &12836 shot texts &pants-fire/ false /barely-true /half-true /mostly-true /true &- &text & {It collected a decade-long, 12.8K manually labeled short statements in various contexts from POLITIFACT.COM, which provides detailed analysis report and links to source documents for each case.}  \\

Twitter & \citet{24} & 13924 news &fake/real &7898/6026 &text, image &{This dataset contains 13924 news and 514 imgaes from MediaEval. It's \textbf{imbalanced dataset}} .\\

WeChat Datasets& \citet{94} &27161 articles &fake/real &2090/2090 &text &{This dataset collected from WeChat’s Official Accounts, dated from March 2018 to October, 2018.It contains 27161with 37971 reports,and 22981 articles is unlabeled. It's \textbf{balanced dataset.} }\\

Twitter & \citet{97} &4 million Tweets &fake/real &- &text &{This dataset is unpublished and contains 4 million tweets, 3 million users, 28893 hashtags, and 305115 linked articles, revolving around 1022 rumours from 01/05/2017 to 01/11/2017.}\\

Twitter & \citet{109} &126000 rumor cascades &rumor/non-rumor & - &text &{126000 rumor cascades spread by 3 million people more than 4.5 million times from Twitter  inception in 2006 to 2017.}\\
 
Weibo-20 &  \citet{57}& 6362 articles  &fake/real &3161/3201 &text &{This dataset is constructed on thr basis of  Weibo-16( \citet{19}), it contains 6362 articles whit 1983440 comments. It's \textbf{balanced dataset.}}\\

FEVER &  \citet{103} &185,446 statements &support /refuted /not enough info &93367/43107/ 48971 &text &{It contains judgment and evidence of the statement.And the statements are from Wikipedia. It's \textbf{imbalanced dataset.}}\\

Fakeddit &{\citet{102}} & 825100 news & fake/real; completely true /fake news with true text /fake news with false text ; true /Satire(Parody) /Misleading Content /Imposter Content /False Connection  &- &text, image, user& {Dataset for 2-way , 3-way, and  5-way classification. It sourced from Reddit. It contains 628,501 Fake samples and 527,049 True samples. It also contains 682,996 Multimodal samples and 358,504 users. It's \textbf{imbalanced dataset.}}\\ 

NewsBag& \citet{101} &215000 news &fake/real &200000 /15000 &text, image&{It contains 200,000 real news , 15,000 fake news and 15000 images. It's \textbf{imbalanced dataset.}}\\

Fake News Evolution & \citet{104} &950 paired data &real/fake /Turned into fake news & - &text &{This dataset includes 950 pieces of data, each of them contains three articles representing the three phases of the evolution process, and they are the truth, the fake news and the evolved fake news.}  \\

\hline

\end{longtable}

\end{center}
\end{sidewaystable}

\section{Potential Issues and Future Work}

{In the past few years, to make the information contained in the network more reliable, researchers worldwide have done much work and made considerable improvements. However, some key areas have not been solved well. In order to answer the RQ5, this chapter focuses on the shortcomings of current research and gives the challenges faced by rumor detection and potential directions for future research.}

{\subsection{Data Ethics Issues}}

{Although the research on rumors detection is vibrant, it has been applied to some scenarios, such as Weibo, rumors have been improved. However, in the process of exploration in this field, some ethical issues have been faced. In the research process, most social science researchers collect Twitter data, such as text content, pictures, videos, and even private information such as the gender of the account owner, the age of the user, and ethnicity. Twitter is particularly challenging in ethics because its data is partially accessible. Although the terms of service stipulate that users’ public posts will be provided to third parties, and by accepting these terms, users will legally agree to this, but surveys (~\citet{2017Towards}) show that less than two-thirds of people have read these terms in whole or in part. Only 76\% of this group know that when they accept the terms of service, they agree to a third party to access some of their information. Although some studies have noticed this problem, they did not disclose their information or hide private information such as user ID for confidentiality or privacy protection, but there is still a problem of privacy leakage. For example, some public rumor data sets (such as Fakeddit(\citet{102}), Twitter Dataset(\citet{24}) have compassionate information related to politics and social figures, but no effective measures have been taken to protect privacy. Furthermore, most of them have not obtained the informed consent of users. In addition, survey data shows that 49\% of respondents from the government and 51\% of commercial companies have concerns about the use of the Twitter data set for research, and respondents expressed a high degree of agreement with the Twitter research on consent to anonymity, and hope to get their consent before publishing academic results using their Twitter posts. Therefore, the issue of data ethics is one of the issues that need to be paid attention to in future research of rumor detection.}

{\subsection{Algorithmic Ethics Issues}}

{\citet{10} stated that the rumor detection algorithm performed well on social media data sets surrounding specific events. However, due to changes in language usage, their accuracy will drop beyond the events they developed. Therefore, as to how researchers should develop, use and reuse algorithms, they usually use sensitive labels to classify content and users without their knowledge, which creates ethical challenges. According to the scale and speed of the data, researchers should ensure that the algorithm performs well on rumor detection to establish standards for rumor classification and ensure that the data labels used are accurate. In addition, if research intends to use rumor classification algorithms for data outside of the designed algorithm data set, the researcher should be responsible for ensuring the continued effectiveness of the classification algorithm because the failure of big data will cause the masses to question the power and longevity of algorithms (\citet{2014The}). Therefore, rumor detection algorithms need to be publicly released and reproduced transparently so that their effectiveness can be tested regularly to avoid mislabeling content and users.}

{\subsection{Legal Issues}}

{The data extracted from the Twitter API contains personal information, which is subject to relevant data protection legislation. Therefore, researchers should establish a fair and legal basis for collecting personal information if the user's informed consent cannot be obtained. Researchers can accept the terms of service of social media networks, and the terms of service provide sufficient conditions to cover relevant data protection legislation. At present, many countries have issued data protection laws, such as the British Data Protection Act (DPA), the Personal Information Protection Law of the People's Republic of China, and other documents (\citet{2017Towards}). Therefore, in the future research process, the collection of relevant personal factor data in social media needs to be collected, used, and disclosed under the relevant data protection law conditions.
}

{\subsection{Real-time Visualization and Learning for Rumor Detection}}

{Although the research of rumor detection is gradually mature, the current rumor detection task is still based on binary classification (Real/Fake) or ternary classification (Real/Fake/Unverified). However, due to the difficulty and complexity of rumor detection tasks in real scenarios, binary or ternary classification models are far from sufficient to identify the characteristics of abnormal online information. Due to the real-time and heterogeneous nature of social communication data, data visualization is a powerful tool to illustrate different aspects and distribution patterns of online social media information. The interactive visualization system can facilitate human supervision and understanding according to the different latitudes and maps of the data, interpret the time-based patterns and behaviors of the data, and summarize important features more clearly. Nowadays, there have been many studies on social media visualization extensively (\citet{201}, \citet{202}, \citet{203},\citet{204}). Except for the work of (\citet{205}), few studies have focused on the interactive exploration of visualizing rumor information on social media. Such a visualization platform can indicate the nature of information dissemination and be regarded as an information resource about the relationship between online users. Through the real-time rumor visualization system, the rumor information or abnormal user behavior can be detected at the moment when the rumor information or user's abnormal behavior occurs. Then an appropriate rumor rejection mechanism can be used to limit the negative impact of the rumor. An online system is a real-time protection measure that attempts to protect online readers immediately.}

{In addition, in combination with real-time rumor detection learning, progressive ideas such as reinforcement learning are used to strengthen the ability of rumor detection and real-time learning of the content and rules of newly erupted rumors. Moreover, there is currently a lack of extensions that deploy research technologies into Web-based real-time verification applications. Such applications should use reinforcement learning methods to learn the characteristics of the latest outbreak of rumors in an emergency and provide the ability to detect rumors. Combined with the real-time rumor visualization system, Internet users or network experts can be one step ahead of the full spread of online rumors, thereby reducing the impact of such information attacks. Usually combined with real-time learning and visualization technology, real-time systems can keep up with the new trend of rumors and can be applied to more scenarios. Therefore, an essential part of online rumor detection and monitoring in real-time visualization systems is an area worth exploring.}

{\subsection{Dataset Issues and  Unsupervised Learning for Rumor Analysis}}

{Although there are many public datasets in the field of rumor detection, and social media such as Weibo and Twitter have opened APIs for scientific researchers to obtain data. However, labeling the rumor detection data set is a tricky thing. Rumors involve many fields, and the work of labeling rumors requires professionals in different fields to complete, and the training of such professionals is challenging and costly. Much money. In addition, even for professionals in the field, labeling news as true or false is a very challenging task for them (\citet{ jahanbakhsh2021semi}). Some works (~\citet{94}) suggest using crowdsourcing to label the rumor detection data set, but the accuracy of the labeling cannot be guaranteed. Some works (\citet{83}, \citet{79}, \citet{150}, \citet{151}) believe that data from some Fact-checking platforms can be used as a rumor detection data set. However, such platforms provide The accuracy of the data is to be investigated, and it is mostly limited to major areas such as politics, medical treatment, and society, and lacks data in proprietary or small domains. Therefore, dataset annotation are difficult, and there is a lack of publicly available large-scale data sets is also one of the challenges to be solved in the future.}

{The unsupervised method can be applied to the actual analysis of real-world data sets. We mentioned the current rumor detection research using unsupervised and semi-supervised learning methods in 5.6. However, we think that in addition to the use of distance measurement (\citet{25}) and methods based on outlier analysis (\citet{27} introduced in 5.6 ), the challenge of the data set can also be solved from the following two directions.}

\vspace{1em}
\begin{enumerate}[i)]
\item {\textbf{Semantic similarity analysis}:  It is used to detect almost repeated news content. Due to the lack of relevant knowledge and imagination of rumor publishers, they often reuse existing rumor content (\citet{dataset1}). For example, a rumor reviewer only needs to modify some essential parts to modify a correct message to be published as a rumor, thereby misleading users. Therefore, semantic similarity analysis to detect tampered rumors can provide a suitable method and can be used for potential rumors detection.} 

\item { \textbf{Unsupervised rumor embedding}: Due to the textual nature of rumors, semantic similarity analysis, sentiment analysis, and other related tasks are essential components of rumor detection. Embedding is an essential step in natural language processing. It refers to the process of extracting high-dimensional representations of original text data. In rumor detection, high-dimensional representations can be used as input for further analysis. Different embedding technologies can capture the characteristics of data from different angles. Choosing a good embedding method plays an essential role in obtaining the underlying nature of news and successfully detecting false information on the Internet. Some popular unsupervised embedding techniques include Word2vec(~\citet{word}), FastText(~\citet{fast}) and Doc2vec(~\citet{doc}).  }

\end{enumerate}

{\subsection{Early Rumor Detection}}

{The current research on rumor detection can only detect rumors when they have been created and have spread to a specific scale through the Internet. However, although some Fact-checking platforms can warn users that the information may contain misinformation, they cannot organize the spread of such rumors in the early stages of the spread of rumors in social media. Furthermore, this kind of problem also needs the attention of future researchers.}

{Therefore, we believe that the early prediction of rumors as future work is significant for rumors detection. Because most of the current rumor detection work is trying to study the true and false detection of the information spread on a large scale, it is essential to detect any trends or potential rumors as early as possible. By learning from historical data, early detection of rumors hopes to find them when the rumors break out. However, there has been work (\citet{early1}) hoping to find the characteristics of rumors in social media in the early stage through the content of the article or the emotional characteristics (\citet{52}). However, there is no guarantee that their generalization capabilities can cope with early rumors detection for newly emerging rumors or new areas of time. Moreover, we believe that early rumors can be detected with more effort, such as potential rumors subject analysis, use of rumor publishers, analysis of data sources, characteristic transfer of different rumors, Etc.}

{\subsection{Rumor Intervention and Refutation Mechanism}}
{Although rumor prediction can remind users of any potential false information during the existence of fake news, the rumor has already caused a certain degree of panic in the process of spreading. However, as far as the current research is concerned, there is still a lack of research on the intervention of rumors and the mechanism of dispelling rumors. (\citet{intervention1}) The influence of rumors in social media has been reduced by constructing a network model for the spread of rumors, and some work (\citet{intervention2}) mentioned some other methods to suppress the spread of rumors, such as deleting the accounts of some high-risk users or making it easier for some people to be vulnerable. Confused users provide immune space. Therefore, the work related to the intervention of rumors is one of the potential research directions to reduce the mass panic and social harm of users caused by the spread of rumors.}

{In addition, the research on the rumor refutation mechanism can understand the mechanism of rumor refutation and rumor intervention ,then disseminate accurate information through a wide range of crucial dissemination nodes. Thereby further reducing the harm of rumors. \citet{refu1} established a dissemination model to study the mechanism of dispelling rumors through the dissemination dynamics of 8 kinds of rumors. \citet{refu2} established a rumor rejection mechanism model based on the principles of game theory. Therefore, the rumor rejection mechanism should also be combined with the intervention of rumors as one of the research directions to reduce the harm of rumors.}

\section{Conclusion}
Rumors and fake news have become by-products of the digital communication ecosystem, and facts have proven very dangerous. Rumor detection work can analyze, characterize, compare and comprehensively evaluate the current scene and classify rumors for the problem. This article attempts to introduce a summary of the latest work in rumor detection. Focus on three aspects: feature extraction, model structure, and methodology. In feature extraction, about 80\% of the research uses machine learning and deep learning to implicitly and explicitly extract rumor features. In the model structure, this article classifies the latest research based on popular structures. The latest work is classified, compared, and introduced according to their respective research directions in the methodology. Finally, the deficiencies of existing research are proposed in the future research for potential research directions for researchers' reference.

To some extent, this work helps new researchers understand the latest developments in the direction of rumor detection and helps newcomers adapt to the field more quickly and sort out their ideas. Fields such as real-time visualization and learning, unsuprevised learning, and early detection of rumors are challenging work that have not yet been resolved and require further research. There is still a long way to improve the accuracy of rumor detection and apply the research content. Overcoming the above challenges will provide further technical support for rumor management and network security.

~\\

\begin{center}
\centering
\footnotesize
\begin{threeparttable}
\begin{tabular}{cl}

\multicolumn{2}{l}{\small{{\textbf{Table 8}}}}\\
\multicolumn{2}{l}{\small{{ The list of abbreviations.}}}\\

\hline
\label{table}\\

\textbf{{Abbreviation}}& \textbf{{Full Name}}\\

\hline
{RQ} & {Research Questions}  \\
{DL} & {Deep Learning}\\
{T} & {Text}\\
{S} & {Social}\\
{V} & {Visual}\\
{PS} & {Propagation Structure}\\
{Acc.} & {Accuracy}\\
{Pre.} & {Precision}\\
{UR} & {Unverified rumor}\\
{TR} & {True rumor}\\
{FR} & {False rumor}\\
{NR} & {Nonrumor}\\
{CNN }& {Convolutional Neural Networks} \\
{TextCNN}& {Text Convolutional Neural Networks} \\
{VGG}& {Visual Geometry Group} \\
{RNN}& {Recurrent Neural Networks}\\
{GRU}& {Gated Recurrent Unit}\\
{Bi-GRU}& {Bi-directional Gated Recurrent Unit}\\
{LSTM}& {Long Short-Term Memory}\\
{Bi-LSTM}& {Bi-directional Long Short-Term Memory}\\
{GNN} & {Graph Neural Networks}\\
{GCN} &{Graph Convolutional Networks}\\
{Bi-GCN}& {Bi-directional Graph Convolutional Networks}\\
{BERT} & {Bi-directional Encoder Representation from Transformers}\\
{TF-IDF}& {Term Frequency–Inverse Document Frequency}\\
{LDA} &{Latent Dirichlet Allocation}\\

\hline

\end{tabular}

\end{threeparttable}
\end{center}

\bibliography{sn-article}









\section*{Statements and Declarations}

There is no interest dispute with this article

\end{document}